\documentclass[graybox]{svmult}
\usepackage{amsmath}

\usepackage{amsmath,amssymb,graphicx,mathtools}
\usepackage{booktabs, multirow}
\usepackage{algorithm,algorithmic,multicol,url,color}

\DeclareMathOperator{\diag}{diag}

\newcommand{\herm}{\mathrm{H}}
\newcommand{\tran}{\mathrm{T}}

\usepackage{array,hhline,subcaption}
\usepackage[cmintegrals]{newtxmath}
 
\usepackage{circuitikz}
\usetikzlibrary{positioning}
\usepackage{tikz}
\usepackage{ellipsis}
\usetikzlibrary{calc}
\usetikzlibrary{decorations.pathreplacing,decorations.markings,shapes.geometric}
\tikzset{naming/.style={align=center,font=\small}}
\tikzset{antenna/.style={insert path={-- coordinate (ant#1) ++(0,0.25) -- +(135:0.25) + (0,0) -- +(45:0.25)}}}
\tikzset{station/.style={naming,draw,shape=dart,shape border rotate=90, minimum width=10mm, minimum height=10mm,outer sep=0pt,inner sep=3pt}}
\tikzset{mobile/.style={naming,draw,shape=rectangle,minimum width=12mm,minimum height=6mm, outer sep=0pt,inner sep=3pt}}
\tikzset{radiation/.style={{decorate,decoration={expanding waves,angle=90,segment length=4pt}}}}

\begin{document}

\title*{Active RISs: Modeling and Optimization}
\author{Recep Akif Tasci\orcidID{0000-0001-9291-0463},\\
Panagiotis Gavriilidis\orcidID{0000-0002-1967-5337},\\
Ertugrul Basar\orcidID{0000-0001-5566-2392}, and\\
George C. Alexandropoulos\orcidID{0000-0002-6587-1371}}
\institute{Recep Akif Tasci \at Department of Electrical and Electronics Engineering, Ko\c{c} University, 34450 Sariyer, Istanbul, Turkey, \email{rtasci20@ku.edu.tr}
\and Panagiotis Gavriilidis \at Department of Informatics and Telecommunications, National and Kapodistrian University of Athens, Panepistimiopolis Ilissia, 16122 Athens, Greece, \email{pangavr@di.uoa.gr}
\and Ertugrul Basar  \at Department of Electrical Engineering, Tampere University, 33720 Tampere, Finland, \email{ ertugrul.basar@tuni.fi}\\ 
Department of Electrical and Electronics Engineering, Ko\c{c} University, 34450 Sariyer, Istanbul, Turkey, email:  and ebasar@ku.edu.tr 
\and George C. Alexandropoulos \at Department of Informatics and Telecommunications, National and Kapodistrian University of Athens, Panepistimiopolis Ilissia, 16122 Athens, Greece, \email{alexandg@di.uoa.gr}
}

\authorrunning{R. A. Tasci, P. Gavriilidis, E. Basar, and G. C. Alexandropoulos}
\maketitle

\abstract{Reconfigurable Intelligent Surfaces (RIS)-empowered communication has emerged as a transformative technology for next generation wireless networks, enabling the programmable shaping of the propagation environment. However, conventional passive RISs are fundamentally limited by the double path loss effect, which severely attenuates the reflected signals. To overcome this, active RIS architectures, capable of amplifying impinging signals, have been proposed. This chapter investigates the modeling, performance analysis, and optimization of active RISs, focusing on two hardware designs: a dual-RIS structure with a single Power Amplifier (PA), and a reflection amplification structure at the unit cell level using tunnel diodes. For the single PA-based design, a comprehensive mathematical model is developed, and closed-form expressions for the received signal-to-noise ratio, bit error probability, and Energy Efficiency (EE) are derived. An optimization framework for configuring the phase shifts and amplifier gain is proposed to maximize system capacity under power constraints. Moreover, a reflection amplification model based on tunnel diodes is introduced, where each unit cell is modeled as a transmission line circuit terminated by a reconfigurable impedance. Regarding the second design, the integration of a tunnel diode into the unit cell is carefully studied by analyzing its I-V characteristic, enabling the derivation of the negative resistance range and the respective power consumption model. Furthermore, the intrinsic phase-amplitude coupling of the reflection coefficient is characterized through compact linear algebra formulations, enabling practical optimization of active RISs. Extensive numerical simulations for both considered architectures validate the theoretical analyses, demonstrating that active RISs can effectively overcome the double path loss limitation, significantly enhance system performance, and achieve favorable EE trade-offs compared to passive RISs, albeit with increased hardware complexity and power consumption. Finally, the fundamental trade-off between the available power budget and the number of active elements is examined, revealing that a higher number of active elements does not always lead to optimal performance.}

\section{Introduction}
The demand for wireless connectivity is growing exponentially, driven by the need to support sophisticated applications~\cite{Marco2019,Saad_6G_2020}, such as digital twinning~\cite{Masaracchia_DT_2023}, automated transportation~\cite{10745876}, urban security~\cite{10230036}, user localization~\cite{Keykhosravi2022infeasible,8313072}, and RF imaging~\cite{Gong_HMIMO_2023,RFimaging}. However, meeting these demands necessitates a paradigm shift in wireless communications, with sixth Generation (6G) communication networks focusing on ultra-reliable, ultra-high-capacity, and energy-efficient, sustainable solutions. Two key enablers of this transformation are eXtremely Large Multiple-Input Multiple-Output (XL-MIMO) systems~\cite{XLMIMO_tutorial_all,shlezinger2021dynamic} and Integrated Sensing And Communications (ISAC)~\cite{10872862,chepuri2023integratedsensingriss}, with a special emphasis on the consideration of the Frequency Range 3 (FR3) band ranging from $7.125$ GHz to $24.25$ GHz. Higher frequencies and larger antenna arrays offer increased range and angular resolution when it comes to localization and sensing, whereas XL-MIMO also offers unprecedented spatial multiplexing gains. However, as the operating frequency increases challenges like increased path loss arise, with the latter necessitating ultra-directive, high-gain beamforming, which is achievable via XL-MIMO arrays. Despite these benefits, the significant increase in antenna elements raises concerns about power consumption, posing a barrier to scaling the number of antenna elements to extreme levels.


Wireless communications empowered by Reconfigurable Intelligent Surfaces (RISs) are becoming a cornerstone of smart and sustainable wireless connectivity towards 6G and beyond networks~\cite{energyeff,basar2019wireless,Marco2019,WavePropTCCN,wu2020intelligent,alexandropoulos2020phaselearning_all,9627818,CE_overview_2022,Alexandropoulos2022Pervasive,RIS_challenges_all}. Unlike traditional massive antenna arrays, which rely on multiple Radio Frequency (RF) chains, RISs consist of nearly passive reflecting elements that manipulate incoming signals by dynamically adjusting their phases~\cite{RIS_Scattering_all}. This reconfiguration is achieved by altering the electromagnetic properties of the reflecting elements, such as their effective permittivity~\cite{9856592}, under the control of a software-defined processor. This enables RISs to transform the wireless propagation environment into a configurable medium~\cite{9673796}, while maintaining minimal power consumption~\cite{ghaneizadeh2024energyharvesting}. As a result, the RIS technology facilitates the development of energy-efficient and cost-effective communication systems~\cite{basar2024risemerging}. Specifically, RISs offer various advantages, including enhanced wireless channel capacity~\cite{9693982}, expanded signal coverage~\cite{RIS_NC}, higher resolution RF sensing~\cite{RIS_smart_cities}, and reduced hardware complexity. Recently, it has been experimentally demonstrated that RISs can improve coverage in THz indoor scenarios~\cite{10701743,10501019}, establishing a strong communication link under direct channel blockage \cite{alexandropoulos2024characterizationindoorrisassistedchannels}. Additionally, RISs mitigate critical challenges in dynamic wireless environments, such as multipath fading and the Doppler effect \cite{yildirim2020modeling,Basar_Doppler,8910627}. 

Due to their outstanding potential, numerous studies have explored the integration of RISs with diverse wireless communication technologies. Notable examples include RIS-based index modulation~\cite{basar2020reconfigurable,10247149}, RIS-enabled reflection modulation~\cite{Miaowen2021,9405433}, and the implementation of Non-Orthogonal Multiple Access (NOMA) with RISs~\cite{9140006,khaleel2020novel,10670007}. Additional research has investigated their use for physical-layer security~\cite{PLS_Kostas_all,alexandropoulos2023secrecyriss,RIS_Privacy}, as well as in sensing-assisted applications, such as user tracking and beamforming~\cite{ghazalian2024jointuserhybridris}, area-wide sensing~\cite{gavras2024simultaneouscommunicationssensinghybrid}, and ISAC~\cite{chepuri2023integratedsensingriss}. Other efforts have targeted waveform optimization in doubly dispersive channels~\cite{ranasinghe2025doublydispersivemimochannelmodel}, RIS-assisted RF energy harvesting~\cite{ghaneizadeh2024energyharvesting}, and channel modeling in the presence of RISs~\cite{basar2020simris,basar2020indoor,kilinc2021physical,antonin2024tacitliniearity,10103817,wijekoon2024physicallyconsistentmodelingoptimizationnonlocal}. While RIS-involved research has yielded significant performance gains and laid a strong foundation with notable academic and industrial impact, passive RISs remain fundamentally limited by two major challenges: the double path loss effect~\cite{basar2021present,dunna2020scattermimo}, and their inability to independently acquire channel state information for self-configuration—a process that can incur considerable signaling and computational overhead \cite{alexandropoulos2022hieararchicalNF,basar2024risemerging,10802983,10600711,zhang2023channel_all}.

Motivated by the previously discussed limitations, researchers have investigated the integration of active components into RIS architectures. This research direction can be broadly categorized into two types: the first includes designs that support baseband conversion and digital signal processing without amplifying the impinging signals—primarily addressing the channel estimation and control problem; the second encompasses architectures that enable reflection amplification, which directly mitigates the double path loss effect and enhances signal strength.
Belonging to the first category,~\cite{9053976} proposed a hybrid RIS architecture where a reception RF chain, equipped with a Low-Noise Amplifier (LNA), was integrated into the RIS to enable baseband signal reception and facilitate channel estimation. A similar approach was adopted in~\cite{alexandropoulos2021hybrid}, where a limited number of reception RF chains were used to reduce training overhead while maintaining estimation accuracy. In~\cite{alamzadeh2021reconfigurable}, channel sensing was realized through a modified RIS structure that couples some incident signals via embedded reception RF chains. These hybrid RISs have also been employed in various ISAC use cases in~\cite{gavras2024simultaneouscommunicationssensinghybrid,ghazalian2024jointuserhybridris,ghazalian2024hybrisris,HRIS_ISAC_radarconf2025_1,HRIS_ISAC_radarconf2025_2}, showcasing its innate ability to enhance both communications and sensing performance. An emerging architecture related to these hybrid designs is the Dynamic Metasurface Antenna (DMA)~\cite{shlezinger2021dynamic}, which generalizes the RIS concept to operate as a compact base station or access point~\cite{9847609}. In DMAs, the RF front-end consists of a metasurface panel whose elements perform either transmission or reception, interconnected via microstrip lines arranged in a row- or column-wise fashion. Each microstrip line is connected to a single RF chain, enabling joint analog and digital signal processing, while significantly reducing power consumption and hardware complexity.

The second category of RISs with active components includes designs that incorporate analog amplification without baseband processing. For instance, the authors in~\cite{zhang2021active} introduced a reflection-type amplifier integrated into each RIS element, which was designed and fabricated, and experimental results were presented validating its performance. In~\cite{9377648}, the performance of such fully active RISs was theoretically analyzed, deriving limits that quantify the trade-off between the number of amplifying elements and the total available power. A distinct hybrid architecture was studied in~\cite{nguyen2021hybrid}, where only a subset of RIS elements is connected to RF chains and Power Amplifiers (PAs), forming a hybrid relay-reflecting intelligent surface that balances hardware complexity and performance. 
Furthermore,~\cite{9758764} proposed an alternative design, in which two passive RISs are interconnected via a single amplifier, forming a structure that emulates an active relay~\cite{Duong_AF_relaying,Duong_relay_selection}, but remains entirely in the analog domain. Finally, it is noted that, unless otherwise stated, throughout this chapter the term active RIS refers specifically to RIS architectures with amplifying capabilities.

While the advancements discussed above have significantly improved RIS technologies, there remains substantial room for further enhancement in system performance and energy efficiency. Designing RIS configurations that can mitigate the double path loss issue more effectively while maintaining cost-efficiency is crucial. By leveraging hybrid architectures that integrate active and passive components or utilizing advanced materials and signal processing techniques, such designs could push the boundaries of current capabilities. In this chapter, the focus is on efficient active RIS hardware architectures. In parcticular, the following Section~\ref{pas_vs_act} provides an overview of the active and passive RIS designs, elaborating on their key advantages and disadvantages. Section~\ref{act_ris} delves into two active RIS architectures, one employing a single PA interconnecting two passive RISs and the other including reflection amplification at the unit cell level via tunnel diodes. Finally, Section~\ref{act_RIS_conc} summarizes the key insights and concludes the chapter.

\section{Passive vs. Active RISs}\label{pas_vs_act}
Selecting between passive and active RIS designs requires careful consideration of their advantages and limitations. Passive RISs are known for their affordability and extremely low energy requirements, making them suitable for energy-efficient applications. On the other hand, active RISs offer enhanced performance by overcoming some of the challenges faced in passive designs, though they introduce additional costs and complexities.

Passive RISs operate with minimal power consumption, which is primarily influenced by their phase resolution and the number of reflecting elements~\cite{energyeff,ghaneizadeh2024energyharvesting,RIS_challenges_all}. Experimental studies have shown that, with the proper placement, even for an \(1\)-bit RIS with \(76\) reflecting elements, the system’s Signal-to-Noise Ratio (SNR) can improve by up to \(10\) dB \cite{10278759}. Furthermore, the authors in \cite{stylianopoulos_1bit} proposed a configuration methodology for the \(1\)-bit case, and analytically derived the performance bounds of \(1\)-bit RISs in SISO systems; a maximum of \(6\) dB loss compared to the gain achieved with infinite phase resolution was shown. This makes them attractive for low-power and cost-sensitive deployments. While these advantages are noteworthy, passive RISs have a significant limitation: the double path loss effect. This issue occurs because the signal weakens twice; once when it travels from the transmitter to the RIS, and again when it reflects from the RIS to the receiver. As a result, the performance improvement of passive RISs diminishes unless they are positioned very close to one of the communication nodes \cite{bjornson2020power,Moustakas2023_RIS}. For instance, when the RIS is located far from both terminals, the double attenuation drastically reduces the signal strength, limiting its practical use in many scenarios.

To overcome the double path loss challenge, researchers have explored integrating active components into RIS architectures. Active RISs use elements such as amplifiers or RF chains to boost the signal strength during reflection, effectively compensating for the energy losses. Recent studies have introduced several innovative designs. Fully-connected active RISs incorporate amplifiers for every reflecting element, ensuring maximum signal enhancement, while sub-connected designs group multiple elements to share a single amplifier, reducing power consumption and cost \cite{9568854}. Hybrid RIS architectures, which combine passive elements for energy efficiency with active components for performance gains, have also shown potential in achieving a balance between cost and efficiency \cite{10235893}. Another approach involves using a single power amplifier to strengthen the reflected signal without drastically increasing system complexity and decreasing energy efficiency \cite{9758764}. These advancements demonstrate that active RISs can overcome the limitations of passive designs, while maintaining the flexibility required for various applications.

However, the adoption of active RISs is not without challenges. One of the main concerns is their higher power consumption compared to passive designs. While the amplifiers help compensate for the double path loss, they contribute to the overall energy demand, potentially conflicting with the goal of energy-efficient communication systems \cite{zhang2021active}. Additionally, active RISs introduce greater complexity to the system, requiring more sophisticated hardware, and consequently, software for their operation. This added complexity can make their deployment more difficult and maintenance more expensive. Cost is another significant drawback, as the need for amplifiers, RF chains, and other active components increases the production and implementation costs compared to the simpler, passive counterparts \cite{zhang2021active, bjornson2020power}. While these issues persist, the superior performance of active RISs ensures their suitability for scenarios requiring high performance.

Considering the advancements in RIS technology, it is clear that incorporating active elements can significantly enhance system performance and energy efficiency. However, achieving a practical and scalable design requires optimizing the power usage, cost, and complexity of active RISs. Researchers continue to explore ways to strike this balance, developing novel architectures to optimize the trade-off between performance and power usage/complexity. In the following sections, we will delve into some innovative active RIS designs that provide substantial performance improvements while addressing concerns related to energy efficiency and cost.

\section{Active RIS Hardware Architectures}\label{act_ris}

\subsection{Two Passive RISs with a Single Power Amplifier}
An amplifying RIS design that incorporates a single PA and two passive RIS panels is presented hereinafter~\cite{9758764}. According to this design, the signals received by the elements of the first RIS are combined after getting their phase configured. Then, the combined signal is fed to the single PA, which feeds it to the second RIS that transmits it with a controllable phase configuration. This design works solely in the RF domain, similar to the waveguide-based approach in \cite{alamzadeh2021reconfigurable}, unlike full-duplex multi-antenna Decode-and-Forward (DF) relays~\cite{DF_relays,alexandropoulos2020full}, which perform down-conversion and baseband processing~\cite{HybridRIS}. Therefore, the design presented in this section looks more similar to a full-duplex, multi-antenna, and Amplify-and-Forward (AF) relay~\cite{Duong_AF_relaying}. However, they have some key differences. Multi-antenna relays usually perform linear processing techniques, such as Maximum Ratio Combining (MRC), and realize power allocation optimization algorithms. Furthermore, they include bulky networks of phase shifters for transmit beamforming and are subject to loopback self-interference \cite{uysal2009cooperative,9933358}. The design that follows is simpler, does not require complex algorithms, and prevents loopback self-interference via spatial separation in the form of back to back placement of the different RISs. Note that simple self-interference cancellation techniques in the RF domain can be applied~\cite{7997175}. 

A high-level comparison of the architectures for a passive RIS, a fully-connected active RIS, and an AF relay is shown in Table~\ref{compare_table}. Fully-connected active RIS configurations, which integrate an active component with each element, have certain drawbacks, such as increased power consumption and higher costs. These limitations may render fully-connected active RIS designs impractical for large-scale systems due to the large number of active components. In contrast, the design based on a single PA offers advantages in Energy Efficiency (EE) and cost-effectiveness, as it requires only a centralized PA unit.
\begin{table}[!t]
\centering
\setlength{\extrarowheight}{7pt}
\caption{Comparison of a passive RIS, a fully-connected active RIS, and a multi-antenna AF relay}
\label{compare_table}
\resizebox{\textwidth}{!}{%
\begin{tabular}{>{\bfseries}p{120pt}cccc}
\toprule
& \shortstack{\textbf{RIS with} \\ \textbf{Single PA}} & \shortstack{\textbf{Passive} \\ \textbf{RIS}} & \shortstack{\textbf{Fully-Connected} \\ \textbf{Active RIS}}  & \shortstack{\textbf{AF} \\ \textbf{Relay}} \\ 
\midrule
Complexity & Low & Low & High & High \\ 
Power Consumption & Low & Low & High & High \\ 
Interference Cancellation & No interference & No interference & No interference & Signal processing \\ 
Cost & Low & Low & High & High \\ 
Performance & High & Low & High & High \\ 
\bottomrule
\end{tabular}%
}
\end{table}
Moreover, the fully-connected active RIS architecture requires greater complexity, as each active element requires individual gain adjustment. In contrast, this design enables gain control through a single PA, simplifying operation. While fully-connected active RISs can achieve higher performance through amplifiers on each reflecting element, the single-PA dual-RIS design leverages passive beamforming across both the receiving and transmitting sections of the RIS. Compared to passive RISs, the RIS with a single PA can boost system capacity while maintaining EE. The system also counters the double path loss effect through amplification, providing flexibility in positioning the RIS between the transmitter (Tx) and receiver (Rx). Nonetheless, this design introduces additional complexity due to the necessary signal processing tasks, including PA gain tuning.

It is noted that a similar concept was previously explored in \cite{abari2017enabling}, where the received signal is amplified and fed back as leakage to the input. However, that study lacks a detailed mathematical model and does not address system operation optimization. In contrast, the amplifying RIS design presented in the sequel is supported by a comprehensive mathematical model, offering a solid framework for system operation. Additionally, an in-depth analysis of the theoretical Bit Error Probability (BEP), capacity, and EE for RIS with a single PA, benchmarking it against a standard passive RIS structure, will be provided.

\subsubsection{System Model}
\begin{figure}[!t]
	\begin{center}
        \medskip
		\includegraphics[width=0.9\columnwidth]{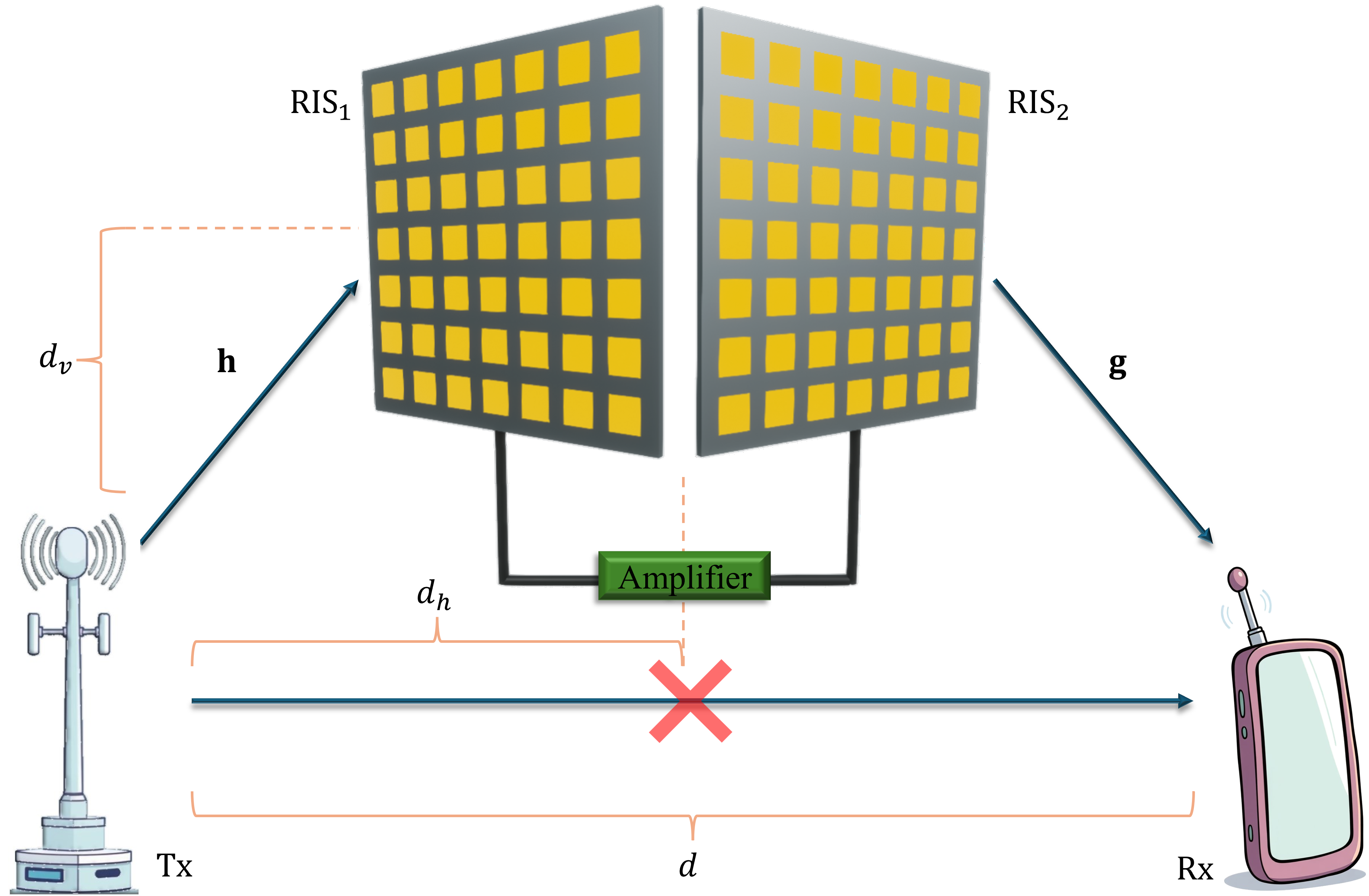}\vspace{10pt}
		\caption{Generic system model including the considered single-PA dual-RIS design enabling the wireless communication of a SISO system.}
		\label{fig:Fig1}
	\end{center} 
\end{figure}

In this section, the amplifying RIS-assisted system model is introduced, using a conventional passive RIS configuration as a benchmark for evaluating the RIS with a single PA. A Single-Input Single-Output (SISO) amplifying RIS-assisted system model with two passive RISs, denoted as RIS\textsubscript{1} and RIS\textsubscript{2}, is considered. Each RIS comprises $N$ reflecting elements, connected via a PA. As shown in Fig. \ref{fig:Fig1}, no direct communication path is assumed between the Tx and Rx. The vertical and horizontal distances between the Tx and RISs, as well as the distance between the Tx and Rx, are represented by $d_v$, $d_h$, and $d$, respectively.

In this model, each element of RIS\textsubscript{1} receives signals from the Tx, which are then combined into a single signal and sent to the PA for amplification. Note that the amplification process also affects the noise added by all elements of RIS\textsubscript{1}. The amplified signal is then transmitted via RIS\textsubscript{2}, where each reflecting element redistributes the signal. To prevent signal leakage from RIS\textsubscript{2} affecting the signals received by RIS\textsubscript{1}, the RISs are positioned back-to-back, as illustrated in Fig. \ref{fig:Fig1}, leveraging the fact that an RIS operates as an one-sided surface. The received complex baseband signal can be mathematically expressed as follows:
\begingroup
\setlength\abovedisplayskip{10pt}
\setlength\belowdisplayskip{10pt}
\begin{equation} \label{eq:1}
y=\sqrt{\frac{G}{N}}\left(\boldsymbol\phiup^\mathrm{T}\mathbf{h}\sqrt{P_t}s+\sqrt{F}n_\text{tot}\right)\boldsymbol\thetaup^\mathrm{T}\mathbf{g}+n_\text{rx},
\end{equation}
\endgroup
where $s$ and $y$ are the transmitted and received signals, respectively; $P_t$ is the Tx power; as well as $G$, $F$, $n_\text{tot}$, and $n_\text{rx}$ are the PA gain, noise figure, total input noise at the PA, and noise sample at the Rx. The vectors $\boldsymbol\phiup \in \mathbb{C}^{N\times 1}$ and $\boldsymbol\thetaup \in \mathbb{C}^{N\times 1}$ represent the phase configurations of RIS\textsubscript{1} and RIS\textsubscript{2}, respectively, with elements $\phi_i$ and $\theta_i$ being the phase shifts of each $i$th element ($i=1,2,\ldots N$).

The channel vectors $\mathbf{h} \in \mathbb{C}^{N \times 1}$ and $\mathbf{g} \in \mathbb{C}^{N \times 1}$ represent the Tx-RIS\textsubscript{1} and RIS\textsubscript{2}-Rx links, respectively. The term $\boldsymbol\phiup^\mathrm{T}\mathbf{h}\sqrt{P_t}s$ denotes the combined signal received by RIS\textsubscript{1}, amplified by $\sqrt{G}$ in the PA, and divided by $\sqrt{N}$ due to the power division in RIS\textsubscript{2}. The amplified and distributed signal is then transmitted to the Rx via RIS\textsubscript{2}. The same scaling applies to $n_\text{tot}$, which is multiplied by $\sqrt{F}$ and $\sqrt{G}$ before being distributed among RIS\textsubscript{2}’s elements. The noise at RIS\textsubscript{2} is ignored, but the noise at RIS\textsubscript{1} is accounted for, since it will be amplified.

Let us assume channels experience Rayleigh or Rician fading, influenced by the probability of Line-of-Sight (LoS), $p_\text{LoS}$, which varies with distance. In particular, the channel coefficients are given by:
\begin{equation} \label{eq:2}
    h_i =  \sqrt{\frac{1}{\lambda^h}} \left(\sqrt{\frac{K_1}{K_1+1}} h^L_i + \sqrt{\frac{1}{K_1+1}}h^{NL}_i\right),\vspace{5pt}
\end{equation}
\begin{equation} \label{eq:3}
    g_i =  \sqrt{\frac{1}{\lambda^g}} \left(\sqrt{\frac{K_2}{K_2+1}} g^L_i + \sqrt{\frac{1}{K_2+1}}g^{NL}_i\right),\vspace{5pt}
\end{equation}
where $K_1$, $K_2$, $\lambda^h$, $\lambda^g$, $h^L_i$, $g^L_i$, $h^{NL}_i$, and $g^{NL}_i$ $\forall i=1,2,\ldots N$ are the Rician factors, path loss, and LoS and Non-LoS (NLOS) components for the elements in the channel vectors $\mathbf{h}$ and $\mathbf{g}$, respectively, with $h^{NL}_i, g^{NL}_i \sim\mathcal{CN}(0,1)$ (i.e., complex Gaussian random variables with zero mean and unit variance). Note that, if there is no LoS, $K_1=K_2=0$ is set. The path losses $\lambda^h$ and $\lambda^g$ depend on whether the link is LoS or NLoS, according to the Indoor Hotspot (InH) environment model from \cite{3GPP_5G}:\vspace{10pt}
\begin{equation} \label{eq:4}
    \lambda_\text{LoS} \text{[dB]}= 32.4+17.3\log_{10}(d_n)+20\log_{10}(f_c),\vspace{5pt}
\end{equation}
\begin{equation} \label{eq:5}
    \lambda_\text{NLoS} \text{[dB]}= \max\left(\lambda_\text{LoS}, 32.4+31.9\log_{10}(d_n)+20\log_{10}(f_c)\right),\vspace{10pt}
\end{equation}
where $d_n \in \{d_1,d_2\}$ corresponds to the Tx-RIS\textsubscript{1} and RIS\textsubscript{2}-Rx distances, $d_1=\sqrt{d_v^2+d_h^2}$ and $d_2=\sqrt{d_v^2+(d-d_h)^2}$, and $f_c$ is the carrier frequency in GHz. The path loss is constant for each $h_i$ and $g_i$ due to the assumption herein that the RIS is in the far field of both the Tx and Rx. For InH, $p_\text{LoS}$ is defined as in \cite{3GPP_5G}:\vspace{5pt}
\begin{equation}\label{eq:6}
p_\text{LoS}=\begin{cases}
1, & d_n \le 5 \\
e^{-\left(\frac{d_n -5}{70.8} \right)}, & 5<d_n \le 49 \\
0.54e^{-\left(\frac{d_n -49}{211.7}\right)}, & 49<d_n
\end{cases}.
\end{equation}
Expanding \eqref{eq:1} to compute the SNR of the amplifying-RIS-assisted SISO system, yields:
\begin{equation} \label{eq:7}
y=\sqrt{\frac{GP_t}{N}}\left(\boldsymbol\phiup^\mathrm{T}\mathbf{h}\right)\left(\boldsymbol\thetaup^\mathrm{T}\mathbf{g}\right)s+\sqrt{\frac{GF}{N}}\left(\boldsymbol\thetaup^\mathrm{T}\mathbf{g}\right)n_\text{tot}+n_\text{rx}.\vspace{5pt}
\end{equation}
The SNR of the RIS with a single PA, using \eqref{eq:7}, is given by:
\begin{equation} \label{eq:8}
\gamma_\text{act}=\dfrac{P_t\left|\sqrt{\dfrac{G}{N}}(\boldsymbol\phiup^\mathrm{T}\mathbf{h})(\boldsymbol\thetaup^\mathrm{T}\mathbf{g})\right|^2}{\left|\sqrt{\dfrac{GF}{N}}\boldsymbol\thetaup^\mathrm{T}\mathbf{g}\right|^2\sigma^2_\text{tot}+\sigma^2_\text{rx}},\vspace{5pt}
\end{equation}
where $\sigma^2_\text{tot}$ and $\sigma^2_\text{rx}$ are the noise powers at the PA input and the Rx, respectively. Hence, the achievable rate for the system is $R_\text{act}=\log_{2}\left(1+\gamma_\text{act}\right)$.

\subsubsection{Performance Analysis}
When aiming to maximize the system capacity with respect to the phase configuration vectors of RIS\textsubscript{1} and RIS\textsubscript{2} as well as the amplifier gain \(G\), the received SNR need to maximized accordingly. Using \eqref{eq:8}, the following optimization problem is formulated:
\begin{align}
        \gamma_\text{act} &= \max_{G, \boldsymbol\phiup, \boldsymbol\thetaup} \quad \dfrac{P_t\left|\sqrt{\dfrac{G}{N}}(\boldsymbol\phiup^\mathrm{T}\mathbf{h})(\boldsymbol\thetaup^\mathrm{T}\mathbf{g})\right|^2}{\left|\sqrt{\dfrac{GF}{N}}\boldsymbol\thetaup^\mathrm{T}\mathbf{g}\right|^2\sigma^2_\text{tot} + \sigma^2_\text{rx}} \nonumber \\
        &\hspace*{0.3cm} \text{s.t.} \quad \left|\phi_i\right| = \left|\theta_i\right| = 1, \quad i \in \{1, 2, \dots, N\}, \nonumber \\
        &\hspace*{1.24 cm} GP_t\left|\boldsymbol\phiup^\mathrm{T}\mathbf{h}\right|^2 \leq P_\text{max}, \nonumber \\
        &\hspace*{1.24 cm} G \leq G_\text{max},\vspace{5pt}
\end{align}
where \( G_\text{max} \) and \( P_\text{max} \) denote the maximum gain of the amplifier and the maximum output power of the amplifier, respectively. In this optimization problem, the optimal value of \( G \) depends on the phase shift matrix of RIS\textsubscript{1}. Therefore, the optimal solution for the phases of the reflecting elements must be determined first. To maximize the received signal power, the signals should be constructively combined at both the PA and the Rx. This can be achieved by adjusting the phase shifts of the channels \( \mathbf{h} \) and \( \mathbf{g} \). The optimal solutions for the reflecting element phases for RIS\textsubscript{1} and RIS\textsubscript{2} follow a constructive combining strategy \cite{8811733}, with phase shifts given by:\vspace{5pt}
\begin{equation} \label{eq:phase1} 
    \phi_i = e^{-j \angle h_i},\vspace{5pt}
\end{equation}
\begin{equation} \label{eq:phase2} 
    \theta_i = e^{-j \angle g_i},\vspace{5pt}
\end{equation}
where \( \angle \cdot \) denotes the phase of a complex number.

In Fig.~\ref{fig:gammas}, the distribution of \( \gamma_\text{act} \) for different values of \(N\) is illustrated: histograms represent the empirical distribution of \(\gamma_\text{act}\) obtained by means of Monte Carlo simulations, while the colored solid lines are Gamma distributions obtained from moment matching with samples of \(\gamma_\text{act}\) for the different values of $N$. It can be observed that the empirical distributions and the fitted Gamma distribution match almost perfectly. Note that the Probability Density Function (PDF) of \(\gamma_\text{act}\) if following the Gamma distribution is given by the well-known expression \cite{lee1979distribution}:
\begin{figure*}[!t]
     \centering\medskip
     \begin{subfigure}[!t]{0.5\textwidth}
         \centering
         \includegraphics[width=\textwidth]{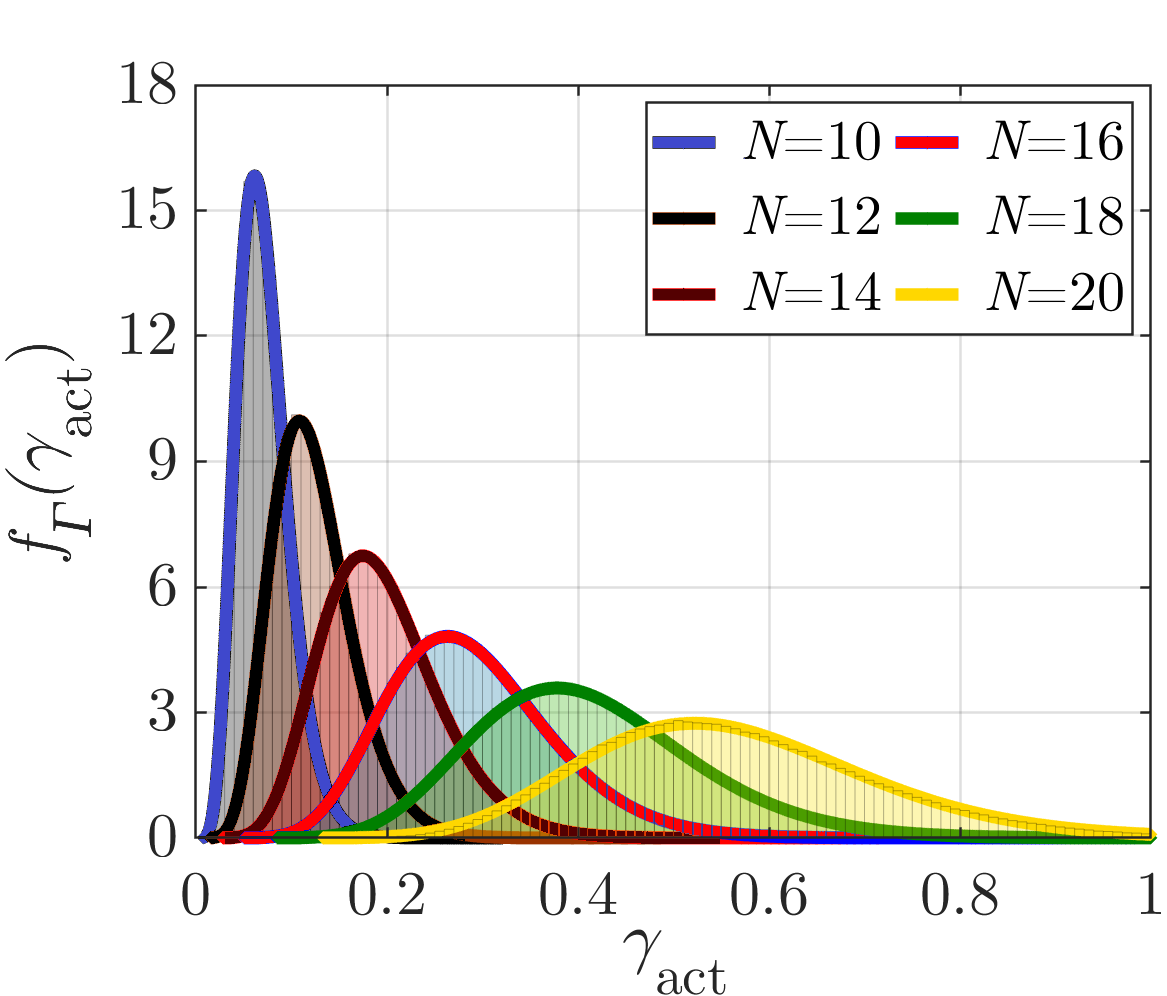}\vspace{10pt}
         \caption{}
         \label{fig:gammasa}
     \end{subfigure}%
     ~
     \begin{subfigure}[!t]{0.5\textwidth}
         \centering
         \includegraphics[width=\textwidth]{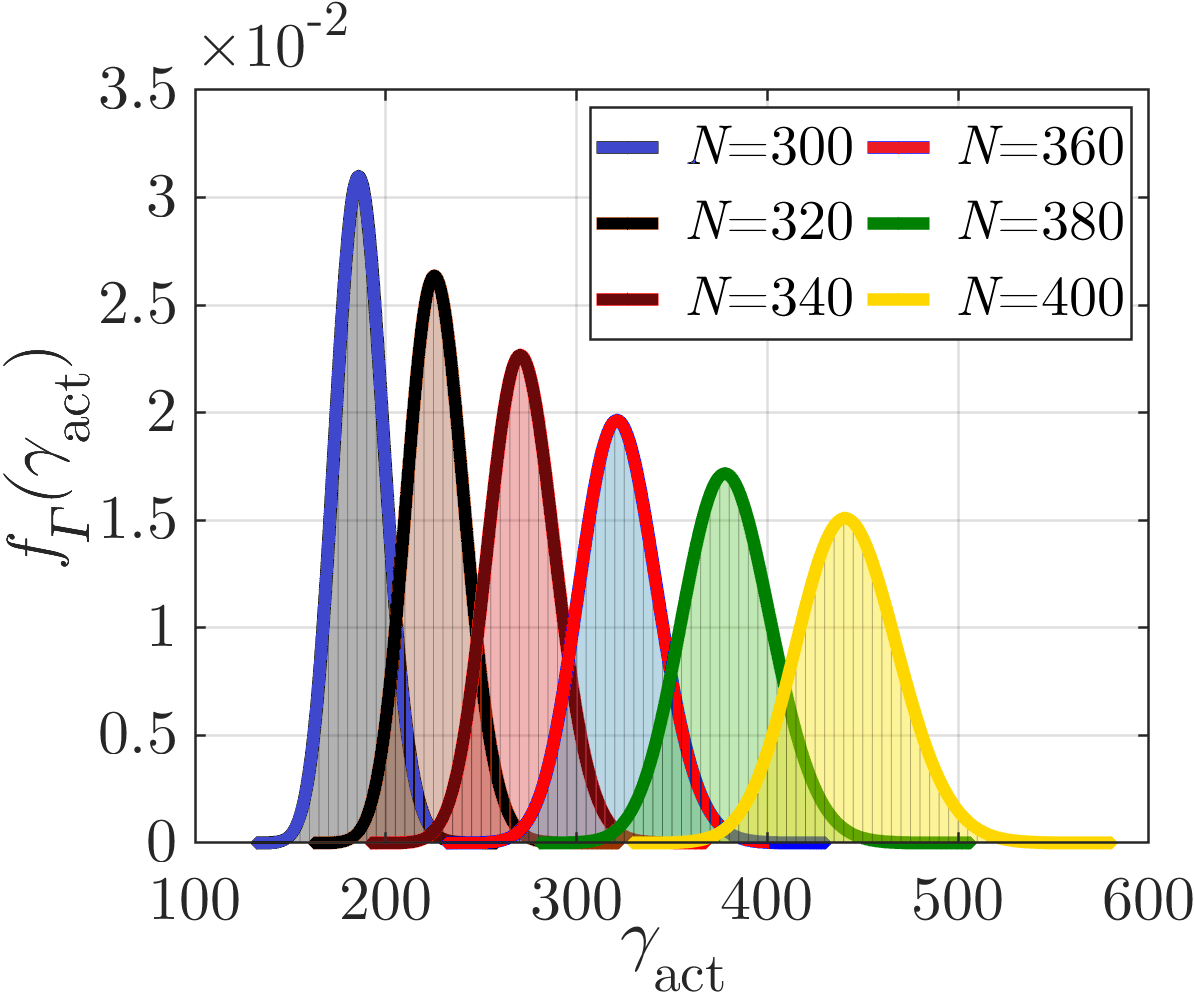}
         \caption{}
         \label{fig:gammasb}
     \end{subfigure}
    \caption{The PDF of $\gamma_\text{act}$ for (a) $N=10,\dots,20$ and (b) $N=300,\dots,400$. The histograms represent the empirical distribution of \(\gamma_\text{act}\) obtained by means of Monte Carlo simulations, while the colored solid lines are the fitted Gamma distributions to \(\gamma_\text{act}\) samples for the different values of $N$.}
    \label{fig:gammas}
\end{figure*}
\begin{equation} \label{eq:Mgf}
f_{\gamma_{\text{act}}}(x) = \frac{{x^{k-1} e^{-x/v}}}{v^k \Gamma(k)},
\end{equation}
where \( \Gamma(\cdot) \), \( k \), and \( \nu \) represent the Gamma function, shape, and scale parameters, respectively. The shape and scale parameters for the fitted Gamma distribution in the \(\gamma_\text{act}\) samples for different values of $P_t$, $P_\text{max}$, and $N$ are provided in Table \ref{Fittable}. It can be seen, for example, that the standard deviation of the distribution increases with \( N \), as also shown in Fig. \ref{fig:gammas}.

\begin{table*}[t]
\setlength{\extrarowheight}{12pt}
\centering
\caption{The shape ($k$) and scale ($\nu$) parameters of the fitted Gamma distribution}
\label{Fittable}
\resizebox{\textwidth}{!}{ 
\begin{tabular}{c|cc|ccccccccc|}
\multicolumn{2}{c}{} &\multicolumn{1}{c}{$P_t\text{(dBm):}$}&-10&-5&0&5&10&15&20&25&\multicolumn{1}{c}{30} \\ \hhline{~===========}
\multicolumn{1}{c|}{\multirow{4}{*}{\rotatebox[origin=c]{90}{$P_\text{{max}}\text{(dBm)}=10\quad\;$}}} & \multicolumn{1}{c|}{\multirow{2}{*}{\rotatebox[origin=c]{90}{$N=64\quad\;$}}}  & $k$        & 44.8922  & 44.7180  & 44.7905  & 44.7109  & 44.8358  & 44.8963  & 48.0934  & 58.6428  & 58.7049 \\ [.5ex]
                                           & \multicolumn{1}{c|}{}                     & $\nu$        & 0.000405 & 0.001287 & 0.004063 & 0.012868 & 0.040595 & 0.128166 & 0.375257 & 0.329909 & 0.329483  \\ [1.5ex] \cline{2-12} 
                                           & \multicolumn{1}{c|}{\multirow{2}{*}{\rotatebox[origin=c]{90}{$N=256\;\;\;$}}} & $k$        & 178.8281 & 178.8481 & 179.1008 & 178.2996 & 233.7027 & 234.8395 & 234.5761 & 233.9432 & 234.3112 \\ [.5ex]
                                           & \multicolumn{1}{c|}{}                     & $\nu$        & 0.006486 & 0.020508 & 0.064762 & 0.205720 & 0.330019 & 0.328426 & 0.328834 & 0.329725 & 0.329212 \\ [1.5ex] \hhline{~===========}
\multicolumn{1}{c|}{\multirow{4}{*}{\rotatebox[origin=c]{90}{$P_\text{{max}}\text{(dBm)}=20\quad\;$}}} & \multicolumn{1}{c|}{\multirow{2}{*}{\rotatebox[origin=c]{90}{$N=64\quad\;$}}}  & $k$        & 44.8284  & 44.7199  & 44.7593  & 44.84389 & 44.8633  & 44.7232  & 44.7599  & 44.7586  & 47.9274  \\ [.5ex]
                                           & \multicolumn{1}{c|}{}                     & $\nu$        & 0.000406 & 0.001286 & 0.004066 & 0.012835 & 0.040551 & 0.128680 & 0.406483 & 1.285651 & 3.765564 \\ [1.5ex] \cline{2-12} 
                                           & \multicolumn{1}{c|}{\multirow{2}{*}{\rotatebox[origin=c]{90}{$N=256\;\;\;$}}} & $k$        & 178.2811 & 178.9068 & 178.4629 & 178.8242 & 178.8931 & 178.6688 & 234.2556 & 234.8709 & 233.5700 \\ [.5ex]
                                           & \multicolumn{1}{c|}{}                     & $\nu$        & 0.006505 & 0.020503 & 0.064994 & 0.205099 & 0.648377 & 2.052960 & 3.292076 & 3.283723 & 3.302386 \\ [1.5ex] \cline{2-12} 

\end{tabular}
}
\end{table*}

Given the previous Gamma distribution fitting, the average Symbol Error Probability (SEP) of the considered amplifying-RIS-assisted SISO system can be analytically evaluated using the Moment Generation Function (MGF) of the Gamma distribution, which is given as follows \cite{arslan2021over}: 
\begin{align}
M_{\gamma_\text{act}}(s) &= (1 - \nu s)^{-k}, \quad \text{for} \quad s < \frac{1}{\nu}.
\end{align}
Using this expression, the average SEP for \( M \)-ary Phase-Shift Keying (\( M \)-PSK) signaling using the MGF in \eqref{eq:Mgf} can be obtained as:
\begin{align} \label{eq:SEP}
P_s &= \frac{1}{\pi} \int_0^{(M-1)\pi/M} M_{\gamma_\text{act}} \left( \frac{-\text{sin}^2(\pi/M)}{\text{sin}^2(x)} \right) dx,
\end{align}
and consequently, the average BEP as \( P_e \approx \dfrac{P_s}{\log_2 M} \). For example, the BEP for binary phase-shift keying (BPSK) simplifies to:
\begin{align}
P_e &= \frac{1}{\pi} \int_0^{\pi/2} \left( 1 + \frac{\nu}{\text{sin}^2(x)} \right)^{-k} dx.\vspace{5pt}
\end{align}

\subsubsection{Total Power Consumption and Energy Efficiency Analysis}
The primary sources of power consumption in the amplifying-RIS-assisted SISO system include the Tx, Rx, PAs (one at the Tx and one between the RISs), and, to a lesser extent, the RIS elements. To start with, for ideal PAs, the power efficiency is given as \cite{amplifier2}:
\begingroup
\abovedisplayskip=15pt
\belowdisplayskip=15pt
\begin{align} \label{eq:power_c}
    \frac{P_\text{out}}{P_\text{amp}} = \eta_\text{max} \left( \frac{P_\text{out}}{P_\text{max}} \right)^{\varepsilon},
\end{align}
\endgroup
where \( P_\text{amp} \) and \( P_\text{out} \) correspond to the power consumed by the amplifier and the output power of the amplifier, respectively. The maximum output power of the amplifier is set to \( P_\text{max} \) to ensure that it operates in the linear region. Here, \( \eta_\text{max} \in (0,1] \) is the maximum efficiency of the amplifier and \( \varepsilon \) is a parameter that depends on the amplifier class. As in \cite{amplifier1}, for more accurate modeling, it is assumed here \( \varepsilon = 0.5 \). The power consumed by the PA can be obtained from \eqref{eq:power_c} as follows:
\begingroup
\abovedisplayskip=10pt
\belowdisplayskip=15pt
\begin{align} \label{eq:pamp}
    P_\text{amp} = \frac{1}{\eta_\text{max}} \sqrt{P_\text{out} P_\text{max}}.
\end{align}
\endgroup

The phase shift of each RIS element is implemented by programmable electronic circuits that consume power as well. The power consumption of the RIS depends on the phase resolution of the RIS elements \cite{energyeffD2D} and is modeled as:
\begingroup
\abovedisplayskip=10pt
\belowdisplayskip=10pt
\begin{align}
    P_\text{RIS} = N P_n(b),
\end{align}
\endgroup
where \( P_n(b) \) is the power consumption of each unit element, which is a function of the phase state resolution. A $6$-bit phase resolution is assumed for each element, leading, according to \cite{energyeff}, to $7.8$~mW power consumption.

Putting all above together, the total power consumption of the  amplifying-RIS-assisted SISO system is given by:
\begingroup
\abovedisplayskip=15pt
\belowdisplayskip=15pt
\begin{align}
    P_\text{tot}^\text{act} = \alpha P_t + P_\text{Tx} + P_\text{Rx} + NP_n(b) + \beta \sqrt{P_\text{out} P_\text{max}},
\end{align}
\endgroup
where \( \alpha = \omega_\text{max}^{-1} \) and \( \beta = \eta_\text{max}^{-1} \), with \( \omega_\text{max} \) and \( \eta_\text{max} \) representing the maximum efficiency of the transmit PA and the PA between the RISs, respectively. For the transmit PA, it is assumed that \( P_t = P_\text{max} \). Here, \( P_\text{Tx} \) and \( P_\text{Rx} \) are the hardware dissipated static powers at Tx and Rx, which are assumed constant and independent of the system parameters. The values for the power consumption model parameters are given in Table \ref{powerpar}~\cite{energyeff}. 
\begin{table}[!t]
\centering
\setlength{\extrarowheight}{3pt}
\caption{Power consumption model parameters}
\label{powerpar}
\begin{tabular}{>{\bfseries}l c}
\toprule
\textbf{Parameter} & \textbf{Value} \\ 
\midrule
$\alpha$           & 1.2      \\ 
$\beta$            & 1.2      \\ 
$P_n(b)$           & 7.8 mW   \\ 
$P_\text{Tx}$      & 9 dBW    \\ 
$P_\text{Rx}$      & 10 dBm   \\ 
\bottomrule
\end{tabular}%
\end{table}

Likewise, the power consumption of the passive RIS-assisted system can be written as follows, only by omitting the power consumed by the PA between the RISs:
\begingroup
\abovedisplayskip=15pt
\belowdisplayskip=15pt
\begin{align}
    P_\text{tot}^\text{pas} = \alpha P_t + P_\text{Tx} + P_\text{Rx} + NP_n(b).
\end{align}
\endgroup

The bit-per-joule EE \( \left( \eta_\text{EE} \right) \) of a system can be computed as:
\begingroup
\abovedisplayskip=5pt
\belowdisplayskip=15pt
\begin{align}
   \eta_\text{EE} = \frac{R_i BW}{P_\text{tot}^i},
\end{align}
\endgroup
where \( BW \) is the communication bandwidth and, for  \( i \in \{ \text{act}, \text{pas} \} \), \( R_i \) is the achievable rate and \( P_\text{tot}^i \) is the total consumed power by the system. learly, the considered here amplifying-RIS-assisted SISO system consumes more power than system using a passive RIS, but provides higher capacity in return. To this end, the EE analysis is an important criterion to determine which system performs better in terms of bits per energy consumption. The EEs under different system configurations are computed in the next subsection through computer simulations without particularly focusing on the optimization of the EE.

\subsubsection{Evaluation of the Error Probability and Achievable Rate}
\begin{figure}[!t]
     \centering
     \begin{subfigure}[!t]{0.7\textwidth}
         \centering
         \includegraphics[width=\textwidth]{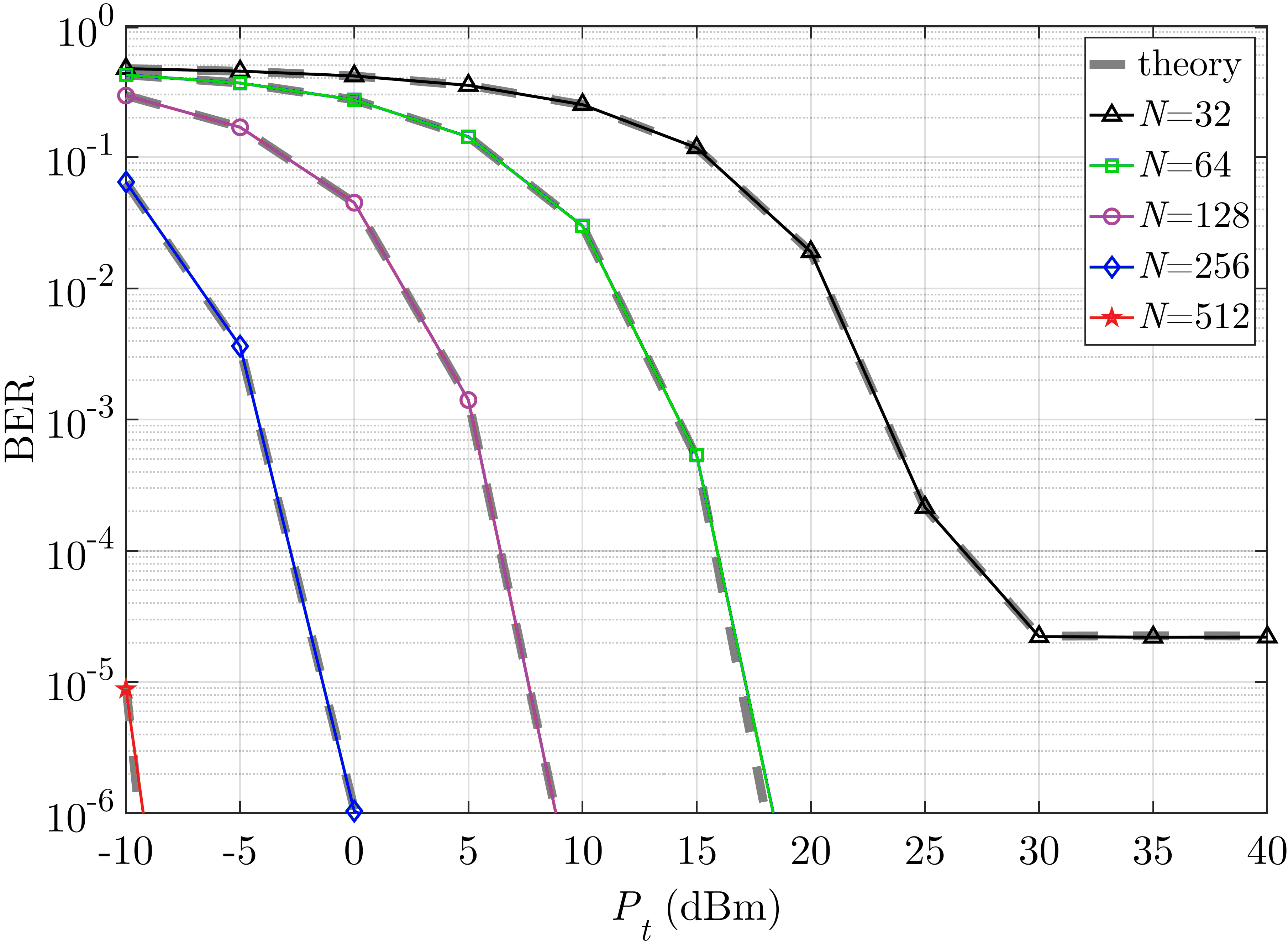}
         \caption{}
         \label{fig:Fig3a}
     \end{subfigure}
     \hfill
     \begin{subfigure}[!t]{0.7\textwidth}
         \centering
         \includegraphics[width=\textwidth]{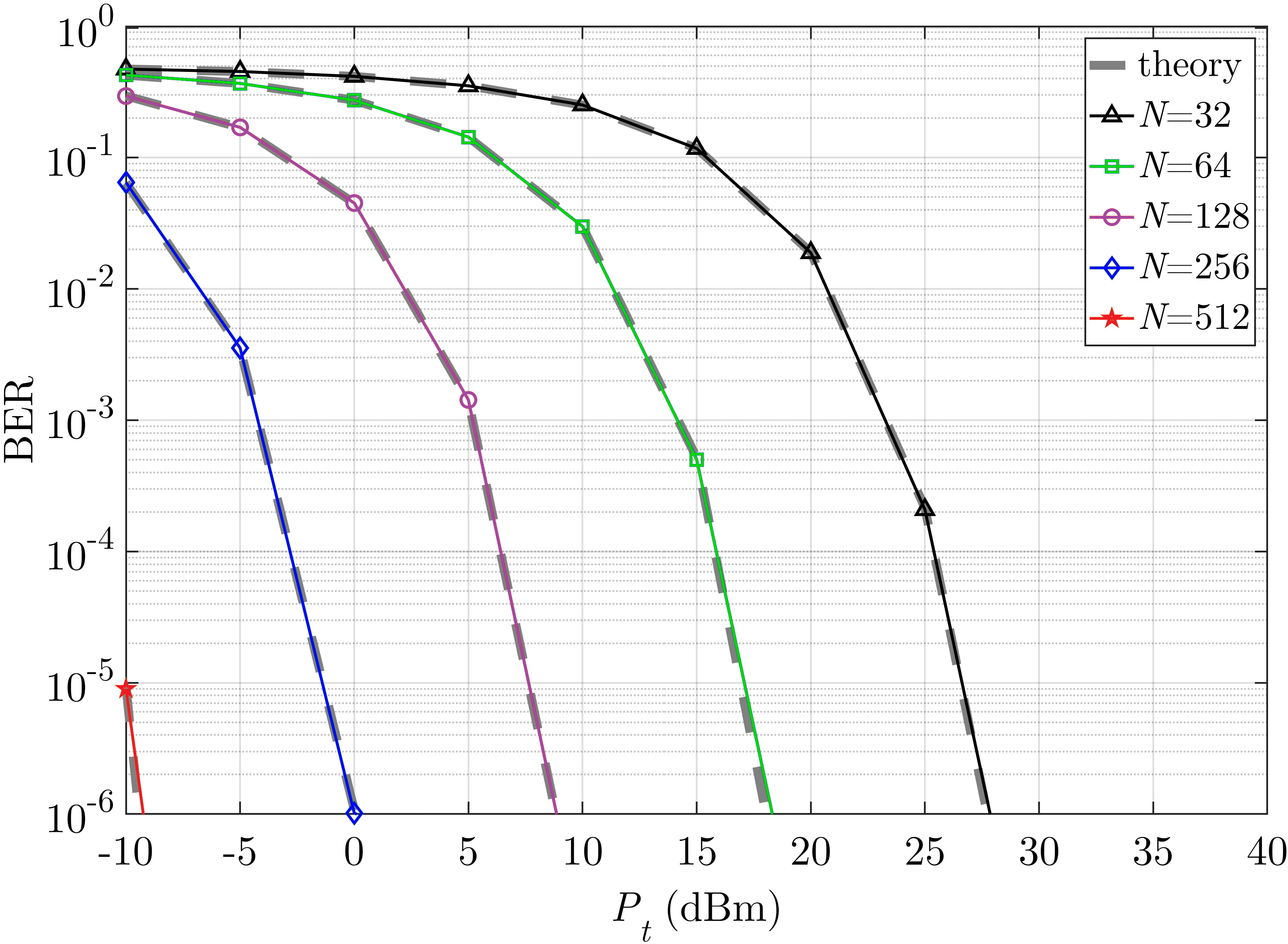}
         \caption{}
         \label{fig:Fig3b}
     \end{subfigure}
    \caption{BER results for (a) $P_\text{max} = 10$ dBm and (b) $P_\text{max} = 20$ dBm.}
    \label{fig:Fig3}
\end{figure}

Consider the system setup of Fig. \ref{fig:Fig1} with the following parameters: $d_v=5$ m, $d_h=5$ m, $d=50$ m, $P_t=30$ dBm, $P_\text{max}=30$ dBm, $f_c=28$ GHz, $BW=180$ kHz, $F=5$ dB, $K_1=K_2=5$, $N=128$, $n_\text{rx}=n_\text{tot}=-100$ dBm, and $G_\text{max}=30$ dB, unless specified otherwise \cite{HMC906A,PA_sample,HMC7054}. For the BER analysis, two different setups were simulated with \(P_\text{max} = 10 \, \text{dBm}\) and \(P_\text{max} = 20 \, \text{dBm}\). Figure~\ref{fig:Fig3a} shows the BER performance for \(P_\text{max} = 10 \, \text{dBm}\), while Fig. \ref{fig:Fig3b} displays the results for \(P_\text{max} = 20 \, \text{dBm}\).
Results from the numerical evaluation of the analytical BER expression (theory) are included together with ones from simulations. It can observed from Fig.~\ref{fig:Fig3a} that an error floor occurs after \(P_t = 25 \, \text{dBm}\) for the \(N = 32\) case. On the other hand, in Fig. \ref{fig:Fig3b}, no error floor is observed. This discrepancy can be explained from the limitation of \(P_\text{out}\), which occurs when the input signal becomes too strong. In this case, the amplifier cannot remain in the linear region if it continues to enhance the signal with the current gain. When \(P_\text{out}\) is fixed to \(P_\text{max}\), further increase in \(P_t\) does not lead to better error performance after a certain point, resulting in the error floor observed in Fig. \ref{fig:Fig3a}. However, in both Figs. \ref{fig:Fig3a} and \ref{fig:Fig3b}, no error floor is observed when \(P_\text{out}\) is smaller than \(P_\text{max}\). Furthermore, a better BER performance is achieved with larger \(N\) and higher \(P_t\) in both cases, as long as \(P_\text{out}\) does not reach \(P_\text{max}\). Based on these observations, it can be concluded that it may be beneficial to keep \(P_\text{max}\) high enough to avoid an error floor. Additionally, it is important to note that, using a larger RIS, can further strengthen the received signal, but an error floor may appear for lower \(P_t\) values due to the limitation of \(P_\text{out}\).

\begin{figure}[!t]
     \centering
     \begin{subfigure}[!t]{0.35\textwidth}
         \centering
         \includegraphics[width=\textwidth]{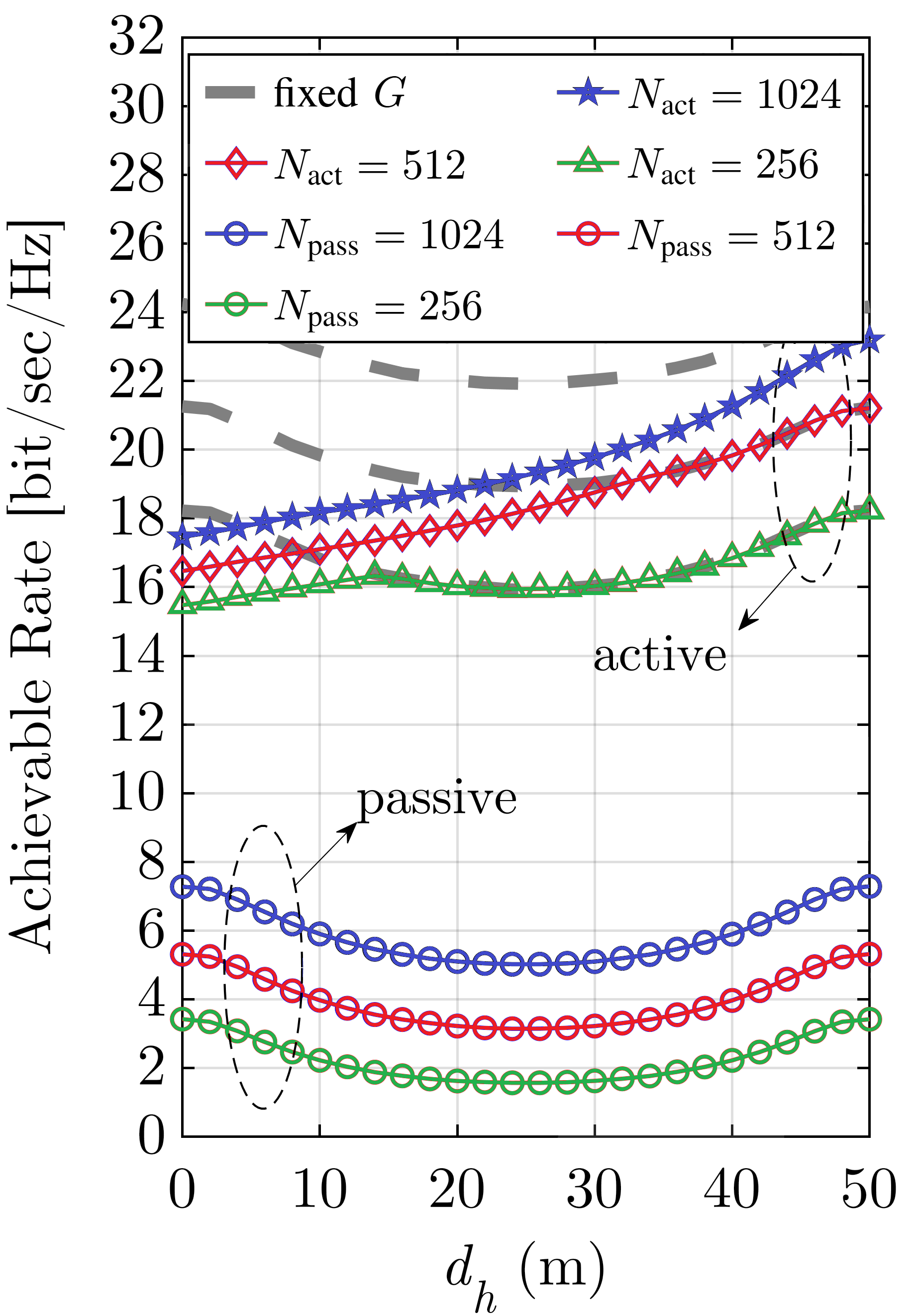}
         \caption{}
         \label{fig:Fig4a}
     \end{subfigure}
     \hspace{30pt}
     \begin{subfigure}[!t]{0.35\textwidth}
         \centering
         \includegraphics[width=\textwidth]{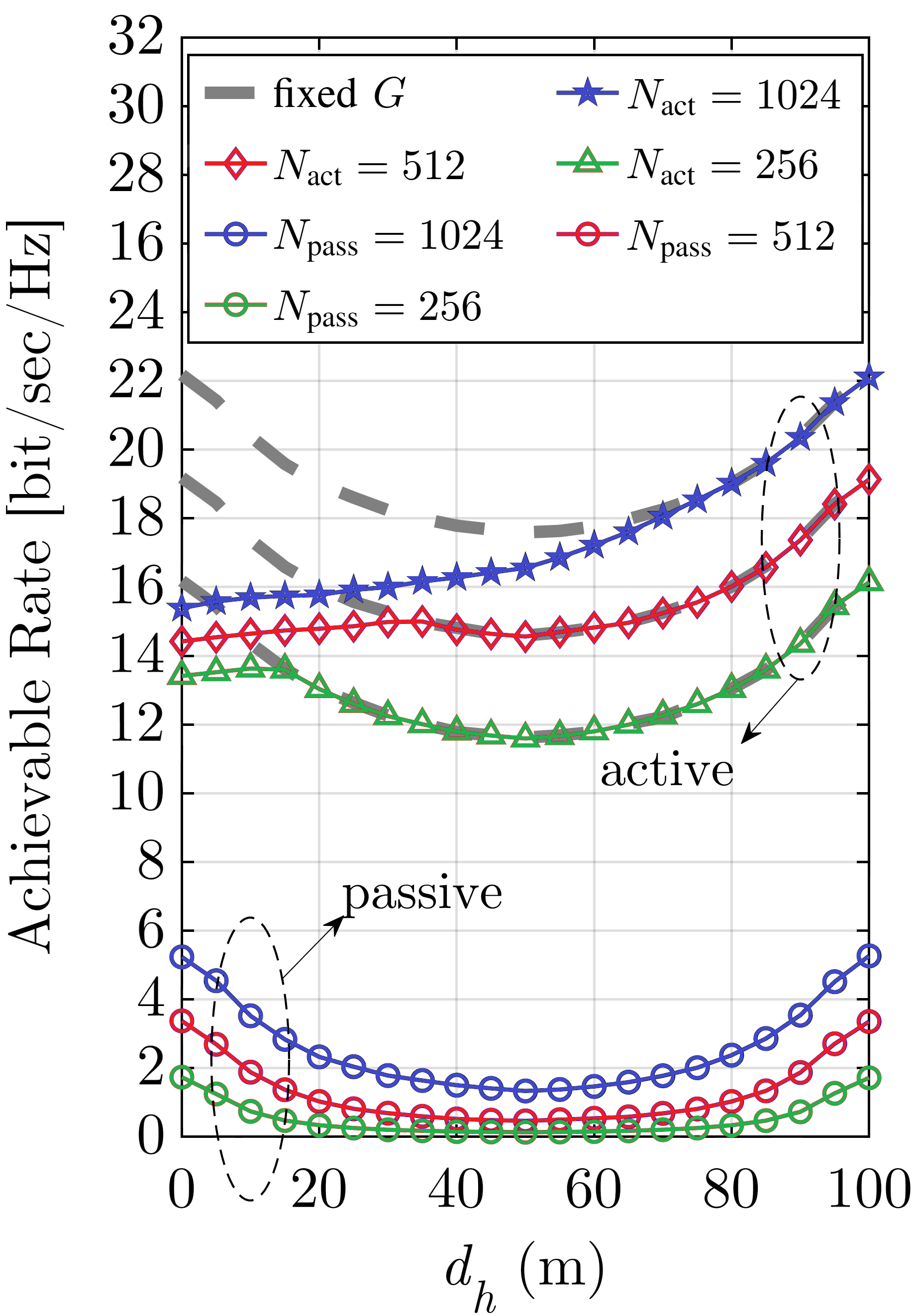}
         \caption{}
         \label{fig:Fig4b}
     \end{subfigure}
    \caption{Achievable rates of the considered amplifying-RIS-assisted SISO system for (a) $d_h=50$ m and (b) $d_h=100$ m. The dashed lines correspond to the case of no output power limitation for the amplifier and \(G\) is fixed at \(30\) dB.}
    \label{fig:Fig4}
\end{figure}

The achievable rates of both designs for varying \(d_h\) with several values of \(N\) are presented in Fig. \ref{fig:Fig4}. Since the proposed active RIS model employs two separate RIS panels, both $\text{RIS}_1$ and $\text{RIS}_2$ contain half the number of reflecting elements compared to the passive RIS. Thus, the total number of reflecting elements in the active design ($N_{\rm act}$) is equal to that of the passive design ($N_{\rm pass}$), ensuring a fair comparison between the two architectures.
The dashed lines show the performance when there is no output power limitation for the amplifier (i.e., results with a fixed \(G\) value of \(30\) dB).
It is shown in Fig.~\ref{fig:Fig4a} that the achievable rates for the active model overlap with the corresponding dashed lines when \(P_\text{out} \leq P_\text{max}\) and \(G_\text{opt} = G_\text{max}\). As it is seen, intersection points with the dashed lines occur at smaller values of \(d_h\) as \(N\) decreases. This behavior arises because \(G_\text{opt}\) starts to decrease after these intersection points, since the amplifying RIS approaches the Tx in order to maintain \(P_\text{out} = P_\text{max}\). Moreover, it is well known that placing an RIS close to either the Tx or Rx generally enhances system performance. This trend is evident in both Figs. \ref{fig:Fig4a} and \ref{fig:Fig4b}, but the amplifying RIS design behaves differently. In this case, placing the amplifying RIS closer to the Tx results in lower achievable rates due to the limitation of \(P_\text{out}\); increasing \(P_\text{max}\) could potentially resolve this issue. Based on these observations, it is concluded that the amplifying RIS has the potential to address a common challenge with passive RIS deployments: the need to position the RIS close to either the Tx or Rx. 

As shown in Fig. \ref{fig:Fig4a}, the passive RIS achieves its lowest rate when positioned midway between the Tx and Rx. This performance worsens in Fig.~\ref{fig:Fig4b} when the Tx and Rx are further apart. The amplifying RIS design significantly mitigates the multiplicative path loss effect by amplifying the combined signal at RIS\textsubscript{1} using a PA between the two RISs. Replacing the passive RIS with an amplifying one, not only offsets this performance drop, but can even enhance it further.

\begin{figure}[!t]
     \centering
     \begin{subfigure}[!t]{0.7\textwidth}
         \centering
         \includegraphics[width=\textwidth]{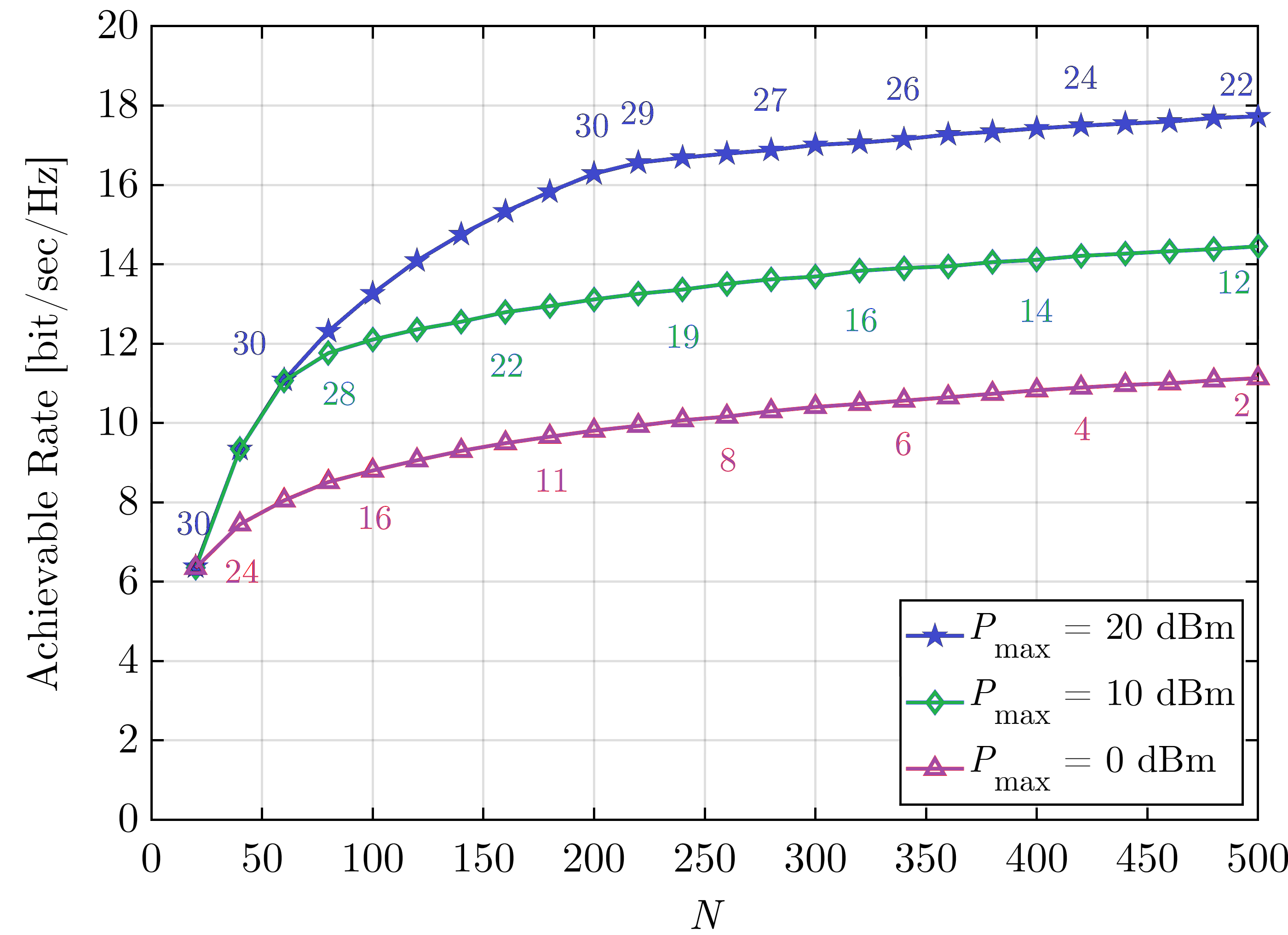}
         \caption{}
         \label{fig:Fig5a}
     \end{subfigure}
     \hfill
     \begin{subfigure}[!t]{0.7\textwidth}
         \centering
         \includegraphics[width=\textwidth]{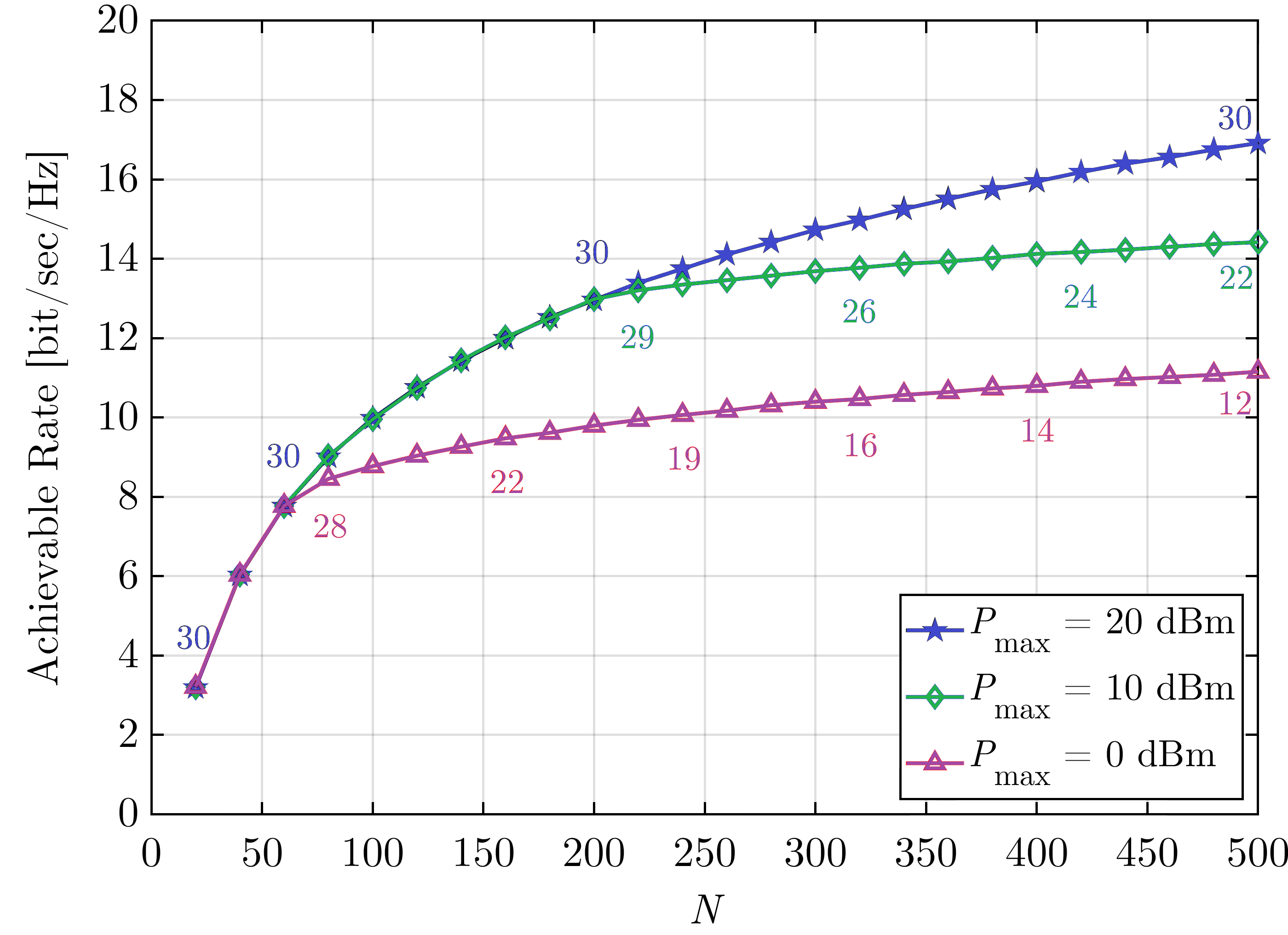}
         \caption{}
         \label{fig:Fig5b}
     \end{subfigure}
    \caption{Achievable rates of the amplifying-RIS-assisted SISO system for (a) $P_t=20$ dBm and (b) $P_t=10$ dBm. The numbers on the markers indicate $G_\text{opt}$ values.}
    \label{fig:Fig5}
\end{figure}

Similar to passive RIS scenarios, the RIS size significantly impacts the achievable rate in active RIS designs as well. Figure \ref{fig:Fig5} illustrates the achievable rates of the amplifying RIS design, with Figs. \ref{fig:Fig5a} and \ref{fig:Fig5b} corresponding to \(P_t = 20\) dBm and \(P_t = 10\) dBm, respectively. Each marker is labeled with its associated \(G_\text{opt}\) for the corresponding \(N\) value. 
In this analysis, it cab be observed that \(G_\text{opt}\) begins to decrease beyond a certain point in all cases except for \(P_\text{max} = 20\) dBm in Fig. \ref{fig:Fig5b}. This behavior differs from the previous case in that, larger \(N\) values lead to an increase in \(P_\text{in}\) as more reflecting elements are employed, which again limits \(G_\text{opt}\). In Fig. \ref{fig:Fig5a}, \(P_\text{out}\) reaches \(P_\text{max}\) for large \(N\), and the amplifier is unable to operate with \(G_\text{max}\) even when \(P_\text{max} = 20\) dBm. However, when \(P_t\) is reduced to 10 dBm, as in Fig. \ref{fig:Fig5b}, \(P_\text{out}\) does not reach \(P_\text{max} = 20\) dBm, even for large \(N\), allowing the amplifier to boost the signal at \(G_\text{max}\). This observation indicates that reducing \(P_t\) in certain configurations allows the amplifier to achieve higher gain levels, provided that \(P_\text{out}\) remains below the maximum threshold.

\begin{figure}[!t]
	\begin{center}
		\includegraphics[width=0.7\columnwidth]{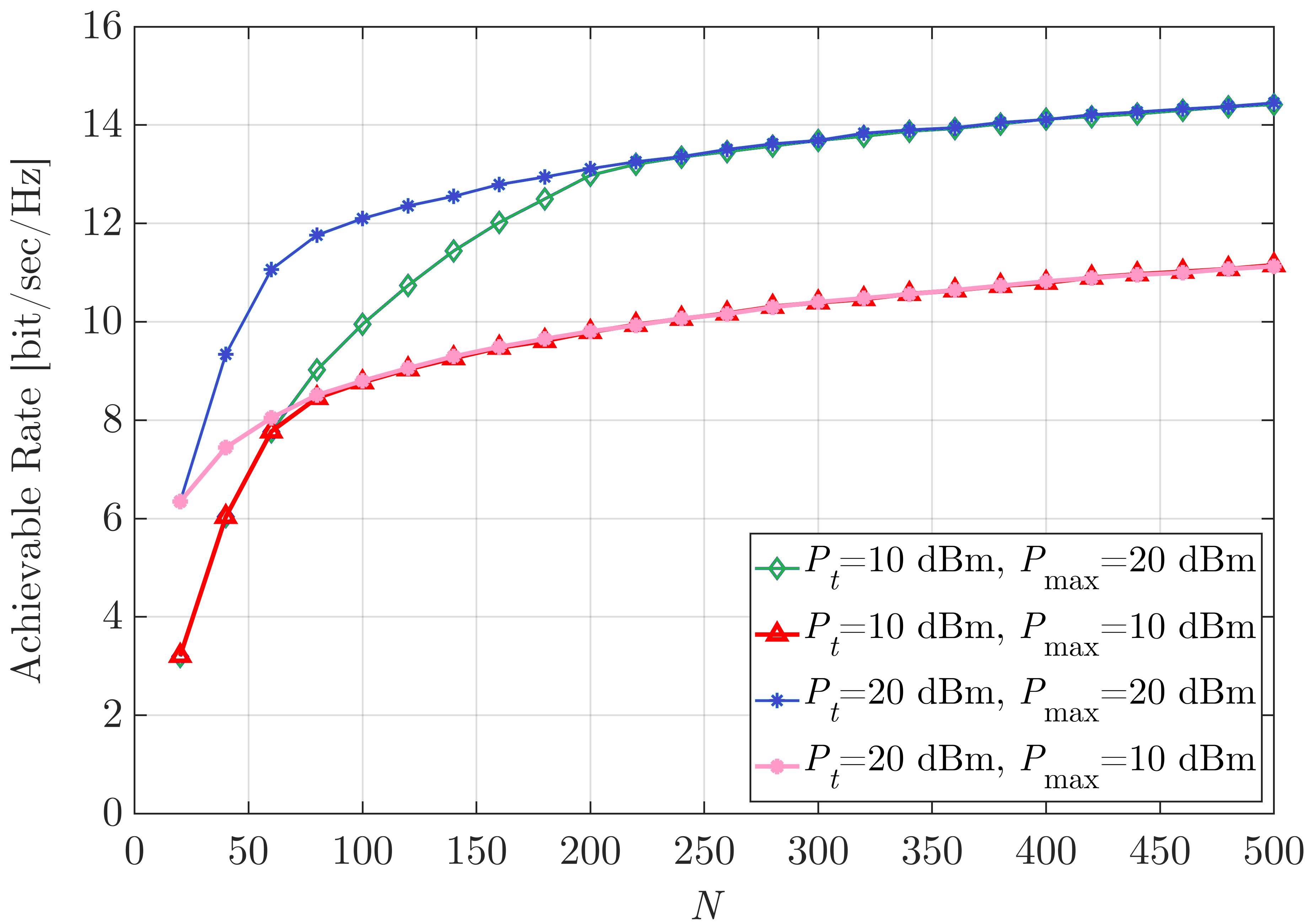}
		\caption{Achievable rates of the amplifying-RIS-assisted SISO system versus the number of elements \(N\) for different $P_t$ and $P_\text{max}$ values.}
		\label{fig:Fig6}
	\end{center} 
\end{figure}

The effects of Tx power (\(P_t\)) and RIS size (\(N\)) on system performance are similar, as both enhance \(P_\text{in}\). In Figs. \ref{fig:Fig5a} and \ref{fig:Fig5a}, it is shown that increasing \(P_t\) does not always lead to additional improvements in system capacity. This phenomenon is further illustrated in Fig. \ref{fig:Fig6}, where achievable rates become identical beyond a certain point for the same \(P_\text{max}\). In Fig. \ref{fig:Fig6}, the achievable rates converge after \(N=200\) and \(N=60\) for \(P_\text{max}=20\) dBm and \(P_\text{max}=10\) dBm, respectively, because \(P_\text{out}\) reaches \(P_\text{max}\) at these \(N\) values. Since \(P_\text{max}=10\) dBm is reached at lower \(P_\text{in}\) values, this intersection occurs at a smaller \(N\) compared to the \(P_\text{max}=20\) dBm case. 
When \(N\) is below 100 for \(P_\text{max}=20\) dBm and below 50 for \(P_\text{max}=10\) dBm, a noticeable performance difference is observed. These results indicate that, increasing \(P_t\), no longer improves performance once the breakpoints are reached. However, the performance can still be enhanced slightly by increasing \(N\). This behavior can be attributed to the two RISs positioned on either side of the amplifier. Increasing \(N\) for RIS\textsubscript{1} raises \(P_\text{in}\) in the same way as increasing \(P_t\). However, if \(P_\text{out}\) has already reached \(P_\text{max}\), further increase in \(N\) for RIS\textsubscript{1} does not impact the achievable rate. On the other hand, increasing \(N\) for RIS\textsubscript{2} does yield slightly higher achievable rates due to the additional beamforming gain associated with a larger \(N\), even though total power remains unchanged and is distributed among more reflecting elements.

All previous results indicate that the primary factors limiting system performance are the PA parameters \(P_\text{max}\) and \(G_\text{max}\). Consequently, the impact of \(N\) and \(P_t\) on performance is highly dependent on them.

\subsubsection{Energy Efficiency Evaluation}
\begin{figure}[!t]
     \centering
     \begin{subfigure}[!t]{0.7\textwidth}
         \centering
         \includegraphics[width=\textwidth]{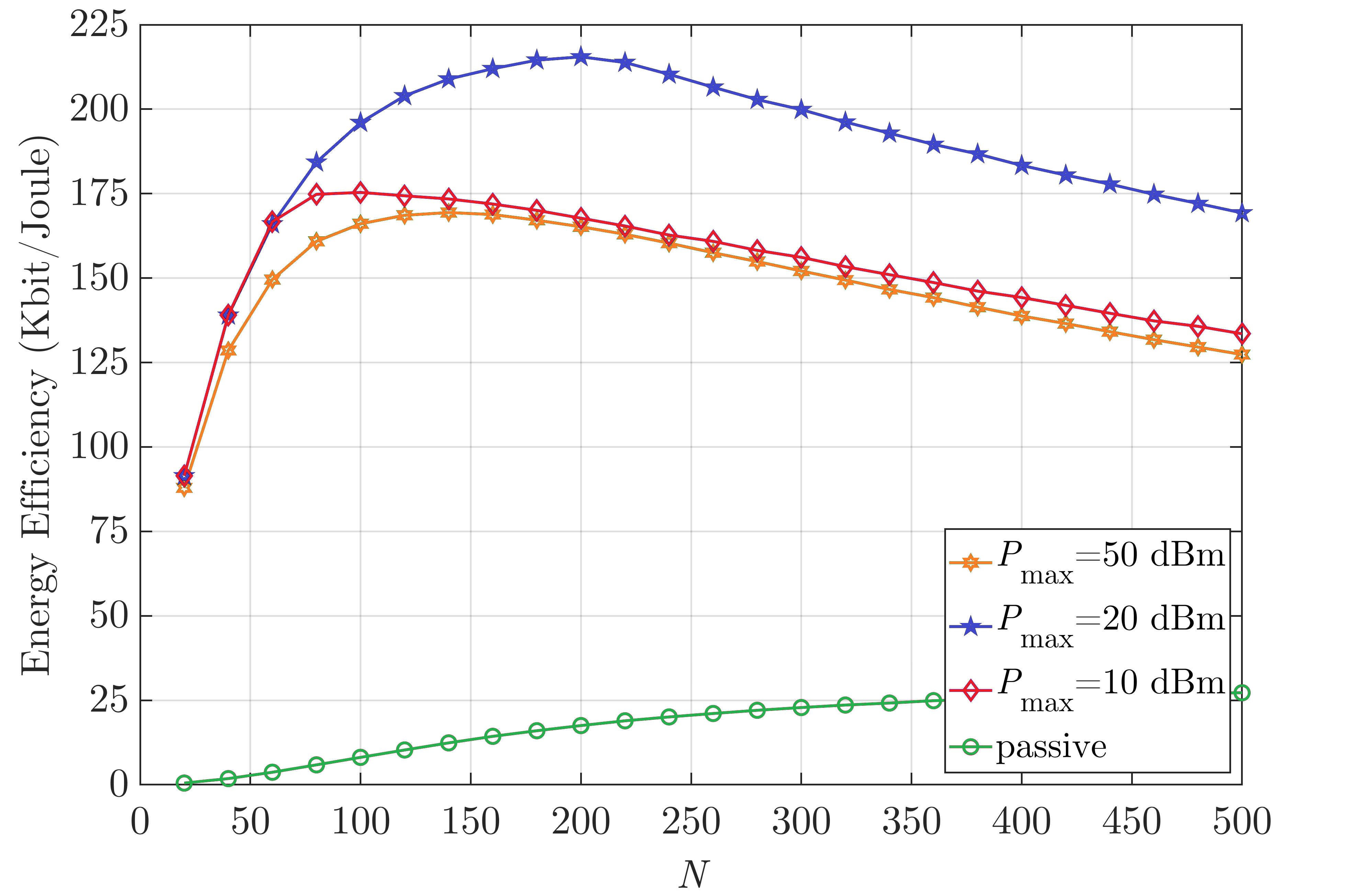}
         \caption{}
         \label{fig:Fig7a}
     \end{subfigure}
     \hfill
     \begin{subfigure}[!t]{0.7\textwidth}
         \centering
         \includegraphics[width=\textwidth]{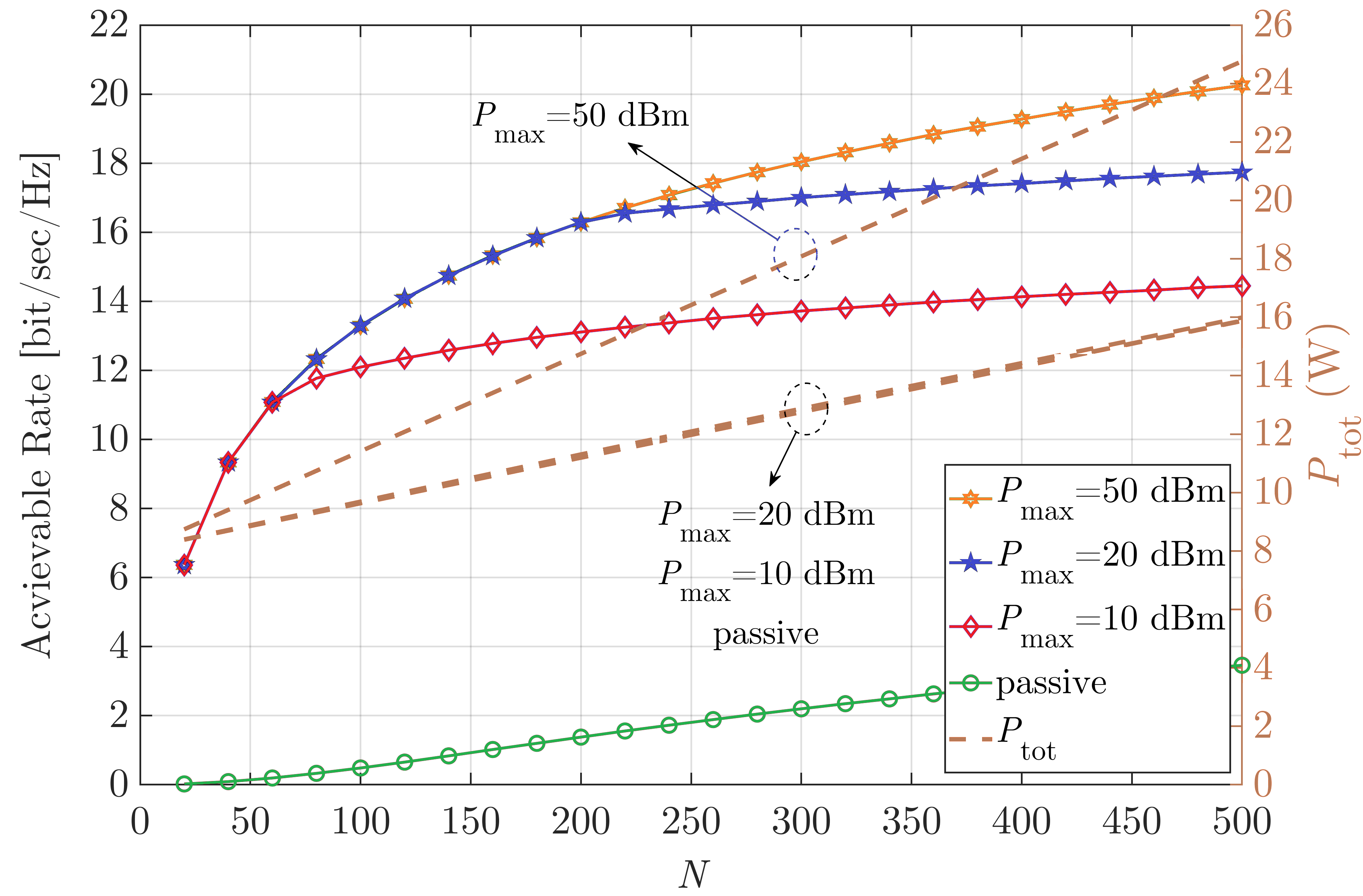}
         \caption{}
         \label{fig:Fig7b}
     \end{subfigure}
    \caption{(a) Energy efficiency of the amplifying-RIS-assisted SISO system versus varying $N$; and (b) corresponding achievable rates and $P_\text{tot}$ values.}
    \label{fig:Fig7}
\end{figure}
\begin{figure}[!t]
     \centering
     \begin{subfigure}[!t]{0.7\textwidth}
         \centering
         \includegraphics[width=\textwidth]{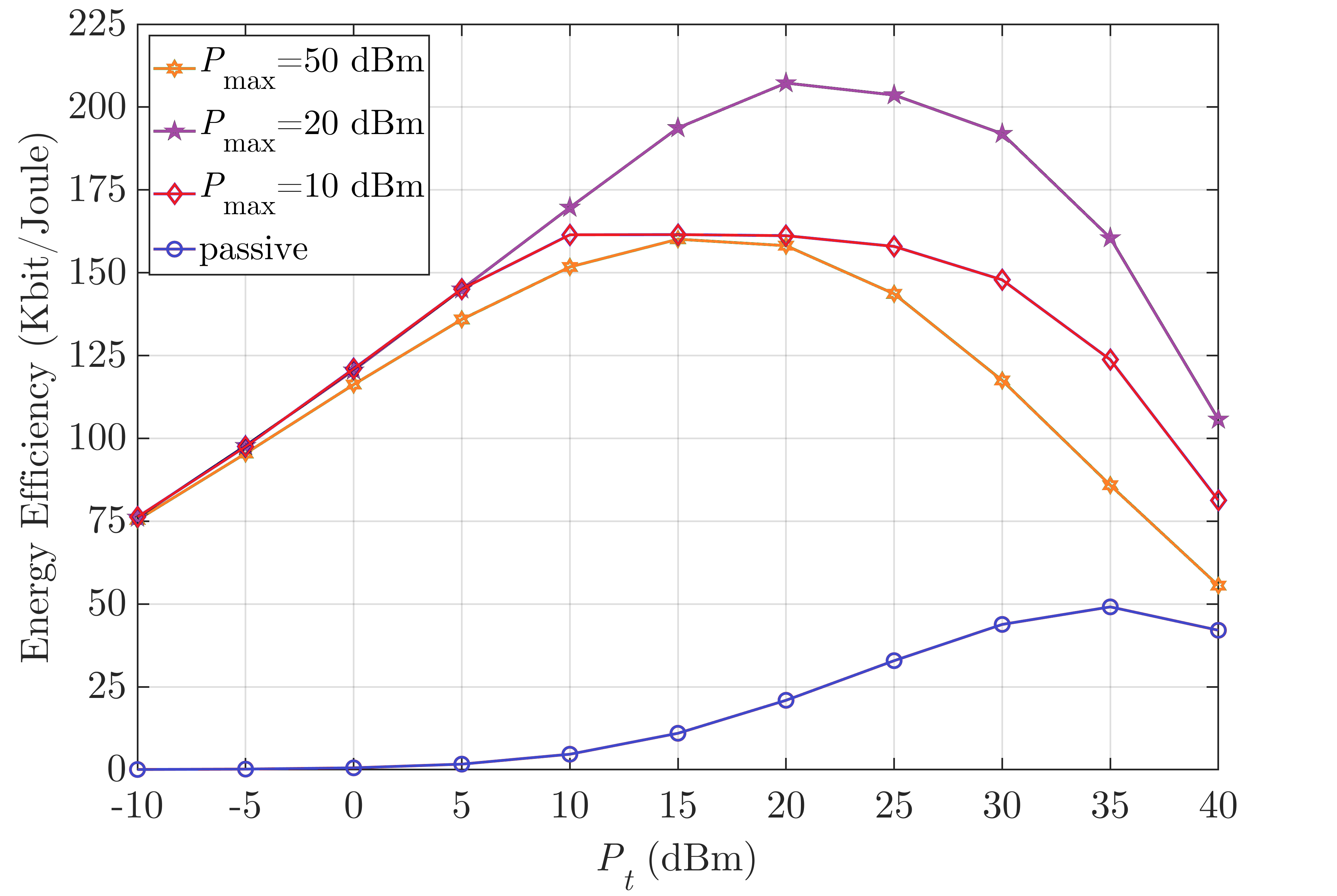}
         \caption{}
         \label{fig:Fig8a}
     \end{subfigure}
     \hfill
     \begin{subfigure}[!t]{0.7\textwidth}
         \centering
         \includegraphics[width=\textwidth]{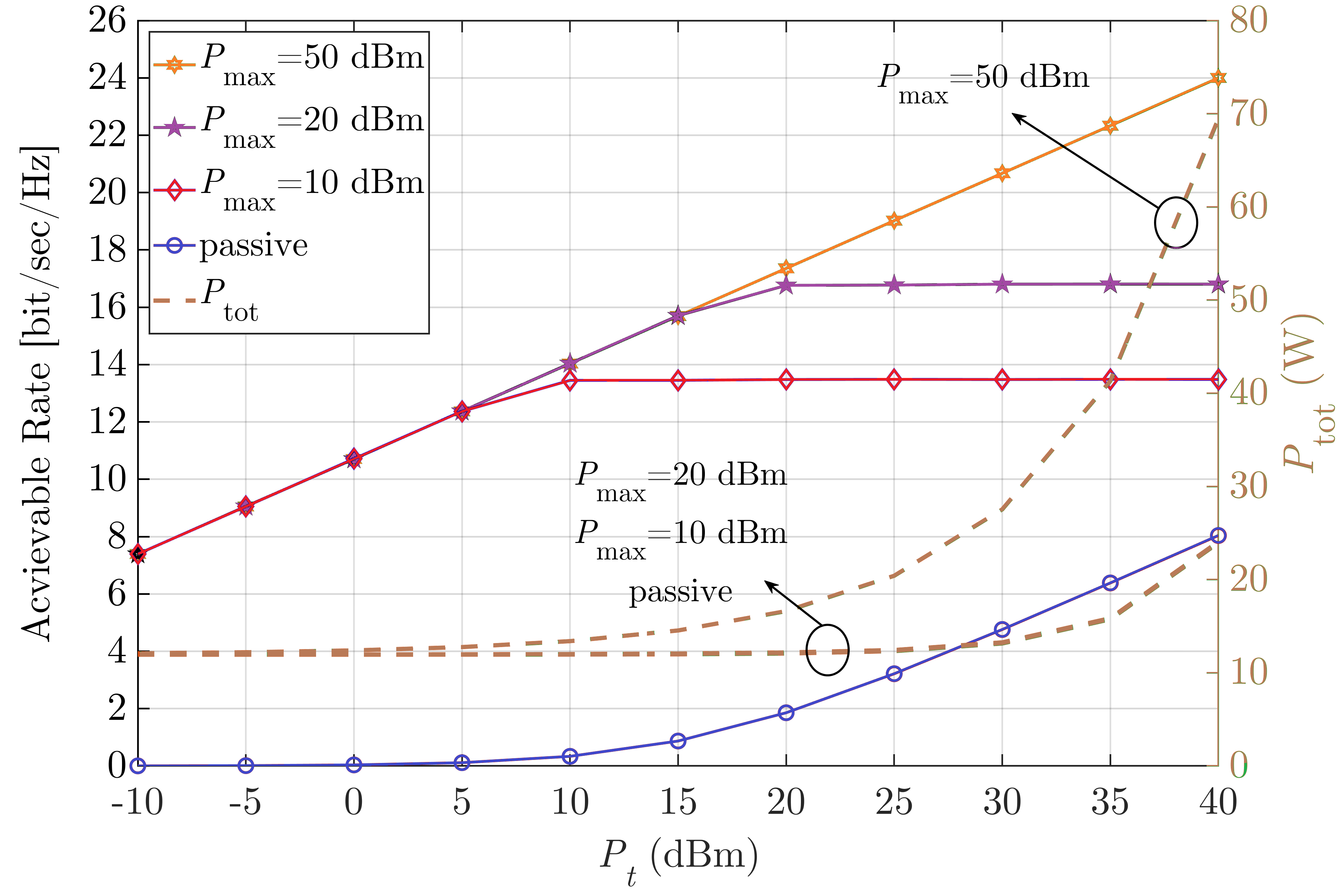}
         \caption{}
         \label{fig:Fig8b}
     \end{subfigure}
    \caption{(a) Energy efficiency of the amplifying-RIS-assisted SISO system for different $P_\text{max}$ values varying with $P_t$; and (b) corresponding achievable rates and $P_\text{tot}$ values.}
    \label{fig:Fig8}
\end{figure}
Figure~\ref{fig:Fig7a} shows the EE values for varying $N$ and different $P_\text{max}$ values, while Fig. \ref{fig:Fig7b} provides the corresponding achievable rates and total power consumption, $P_\text{tot}$. As shown in Fig. \ref{fig:Fig7a}, EE values start to decline after certain points because achievable rates remain nearly constant for $P_\text{max}=20$ dBm and $P_\text{max}=10$ dBm, while $P_\text{tot}$ continues to increase, as illustrated in Fig. \ref{fig:Fig7b}. For $P_\text{max}=50$ dBm, EE performance worsens as power consumption increases, this trend is more pronounced than the achievable rate metrix. However, unless $P_\text{max}$ is exceedingly high, the total power consumption of the active RIS design remains comparable to that of the passive RIS, resulting in better EE for the former.

In Fig. \ref{fig:Fig8a}, EE is evaluated for varying $P_t$ and different $P_\text{max}$ values. Achievable rates and $P_\text{tot}$ values are also provided in Fig. \ref{fig:Fig8b} for the same scenario. It can be seen that, for $P_\text{max}=20$ dBm and $P_\text{max}=10$ dBm, although the amplifying RIS does not consume a substantial amount of power, EE begins to decline after a certain point. This occurs because increasing $P_t$ beyond this threshold, no longer improves achievable rates, yet it still raises $P_\text{tot}$. When $P_\text{max}=50$ dBm and $P_\text{max}$ is not reached, power consumption increases for both the base station’s PA and the PA between the RISs. Consequently, EE decreases beyond a certain point, despite further gains in achievable rate.

\begin{figure}[!t]
     \centering
     \begin{subfigure}[!t]{0.7\textwidth}
         \centering
         \includegraphics[width=\textwidth]{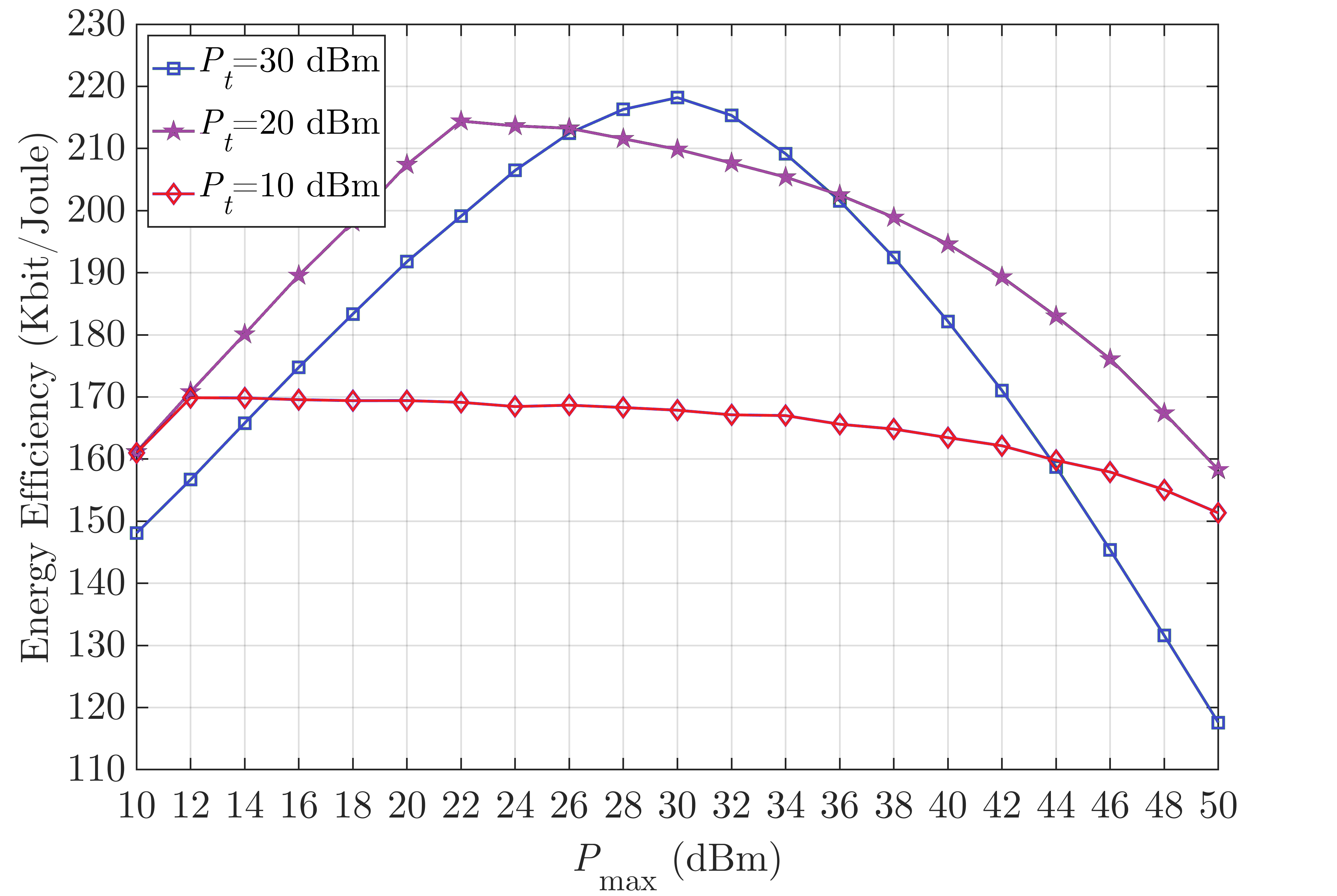}
         \caption{}
         \label{fig:Fig9a}
     \end{subfigure}
     \hfill
     \begin{subfigure}[!t]{0.7\textwidth}
         \centering
         \includegraphics[width=\textwidth]{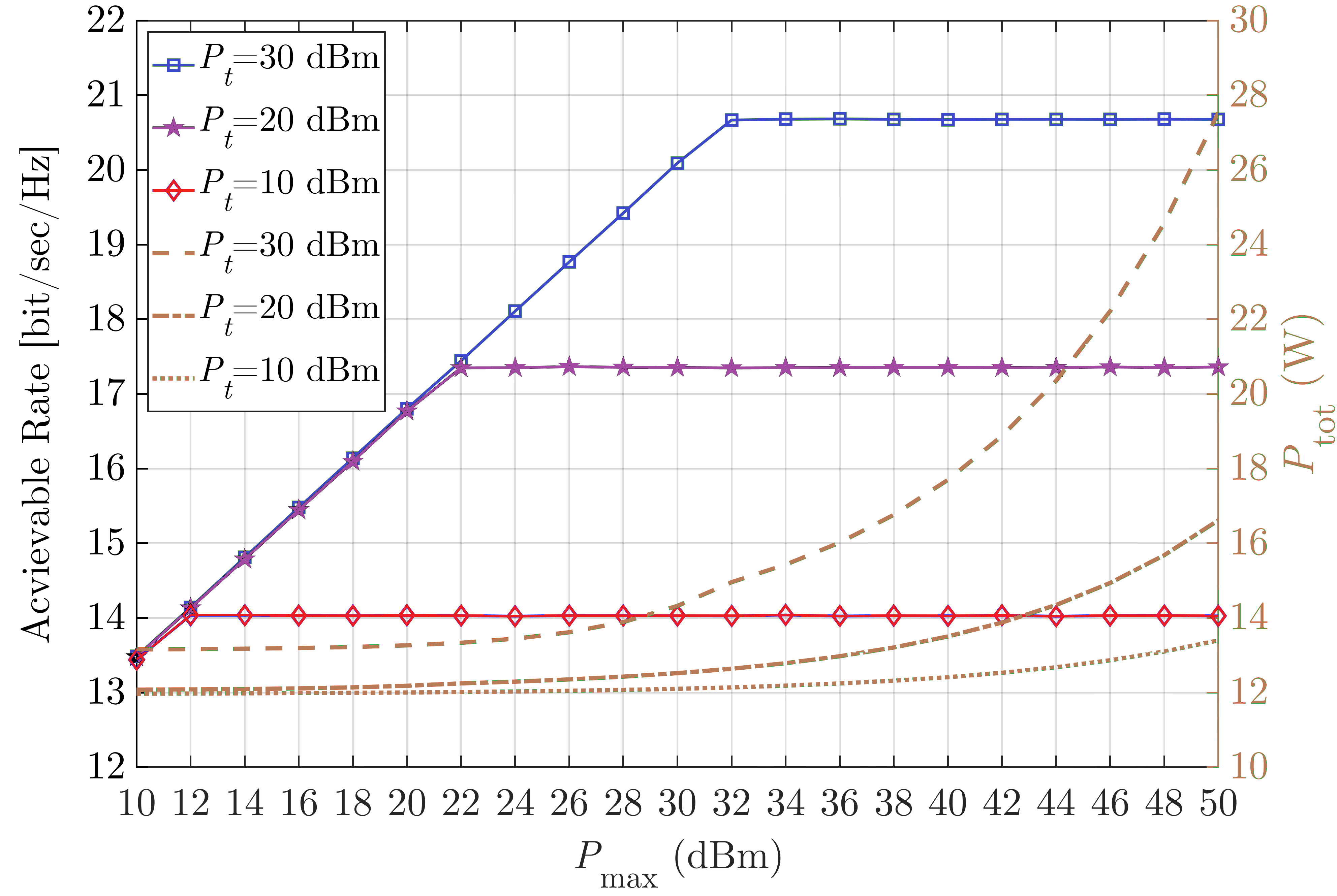}
         \caption{}
         \label{fig:Fig9b}
     \end{subfigure}
    \caption{(a) Energy efficiencies for different $P_t$ values varying with $P_\text{max}$ and (b) corresponding achievable rates and $P_\text{tot}$ values.    
    }
    \label{fig:Fig9}
\end{figure}
Finally, Figs.~\ref{fig:Fig9a} and~\ref{fig:Fig9b} illustrate EE, achievable rate, and $P_\text{tot}$ for varying $P_t$ and different $P_\text{max}$ values. Unlike in Figs. \ref{fig:Fig7} and \ref{fig:Fig8}, the breakpoints in Fig. \ref{fig:Fig9} occur when the amplifier is limited to its maximum gain $G_\text{max}=30$ dB. Increasing $P_\text{max}$ in this case only results in higher power consumption, as described in (\ref{eq:pamp}), without contributing to achievable rate, leading to a downward trend in EE.

All aforedescribed results indicate that EE is influenced by multiple parameters, including $P_t$, $P_\text{max}$, $N$, and $G_\text{max}$. As expressed in \eqref{eq:pamp}, $P_\text{max}$ and $P_t$ have a more direct impact on EE compared to the other parameters. Increasing these values to boost achievable rate does not always result in a more energy efficient system. To maximize EE, the joint optimization of these parameters is required, necessitating advanced optimization techniques; this is a promising direction for future work.

\subsection{Reflection Amplification with Tunnel Diodes}
RISs are typically constructed using planar arrays of passive elements, each capable of dynamically adjusting the phase of reflected EM waves. These elements are often managed through digital control systems implemented on printed circuit boards~\cite{cui2014coding}. By modeling each RIS unit cell as a resonant circuit, it is possible to tailor the reflection characteristics—both amplitude and phase—of incident EM waves through careful adjustment of circuit parameters. By adopting a transmission line circuit model to represent each \(n\)-th RIS element (with \(n = 1, \ldots, N\)), the impedance of this configuration is given as follows~\cite[eq.~(3)]{Abeywickrama_2020}:
\begin{equation}\label{eq:Z_n}
   Z_n(C_n, R_n) \triangleq \frac{\jmath\omega L_1 \left( \jmath\omega L_2 + \frac{1}{\jmath\omega C_n} + R_n \right)}{\jmath\omega L_1 + \left( \jmath\omega L_2 + \frac{1}{\jmath\omega C_n} + R_n \right)},
\end{equation}
where \(L_1\) and \(L_2\) are the inductances associated with the bottom and top layers of the element, respectively; \(\omega\) represents the angular frequency of the incoming signal; and \(C_n\) and \(R_n\) denote the tunable capacitance and resistance, respectively.

Let \(\boldsymbol{\gamma} \in \mathbb{C}^N\) represent the vector of reflection coefficients for an RIS comprising \(N\) elements. The reflection coefficient for each element, \([\boldsymbol{\gamma}]_n\), arises from the impedance mismatch between free space (\(Z_0 = 377\,\Omega\)) and the element's impedance \(Z_n(C_n, R_n)\), and is calculated as:
\begin{equation}\label{eq:ref}
   [\boldsymbol{\gamma}]_n(C_n, R_n) \triangleq \frac{Z_n(C_n, R_n) - Z_0}{Z_n(C_n, R_n) + Z_0}.
\end{equation}
This expression shows clearly that both the magnitude (\(\alpha_n\)) and phase (\(\varphi_n\)) of each RIS reflection coefficient are functions of the adjustable parameters \(C_n\) and \(R_n\). 

\subsubsection{Negative Resistance}
In standard RIS architectures, the circuit parameter \(R_n\) typically represents the resistive losses arising from semiconductors, metallic traces, and dielectric substrates—loss mechanisms that are fundamentally unavoidable in practical implementations~\cite{Gradoni_impedance_modeling,Abeywickrama_2020}. These losses degrade the reflection efficiency, limiting the system’s overall performance. To overcome this limitation, active components can be introduced into the unit cell design, allowing the impedance model in~\eqref{eq:ref} to transition into an amplifying regime. In such configurations, \(R_n\) can have negative values, enabling not only loss compensation but also the realization of controllable reflection gain.

Negative resistance refers to the counterintuitive phenomenon where an increase in voltage across a device results in a decrease in current, effectively reversing the direction of power flow~\cite{amp1}. This behavior can be exploited to amplify incident electromagnetic signals through energy injection from an external DC power source. Among the various devices capable of exhibiting negative resistance, Tunnel Diodes (TDs) stand out due to their compact size, low power requirements, and cost effectiveness. These diodes operate based on quantum tunneling effects in heavily doped p-n junctions, giving rise to a region of negative differential resistance in their current-voltage (I-V) characteristics, which will be elaborated further in the sequel.
\begin{figure}[!t]
    \centering
    \includegraphics[width=\linewidth]{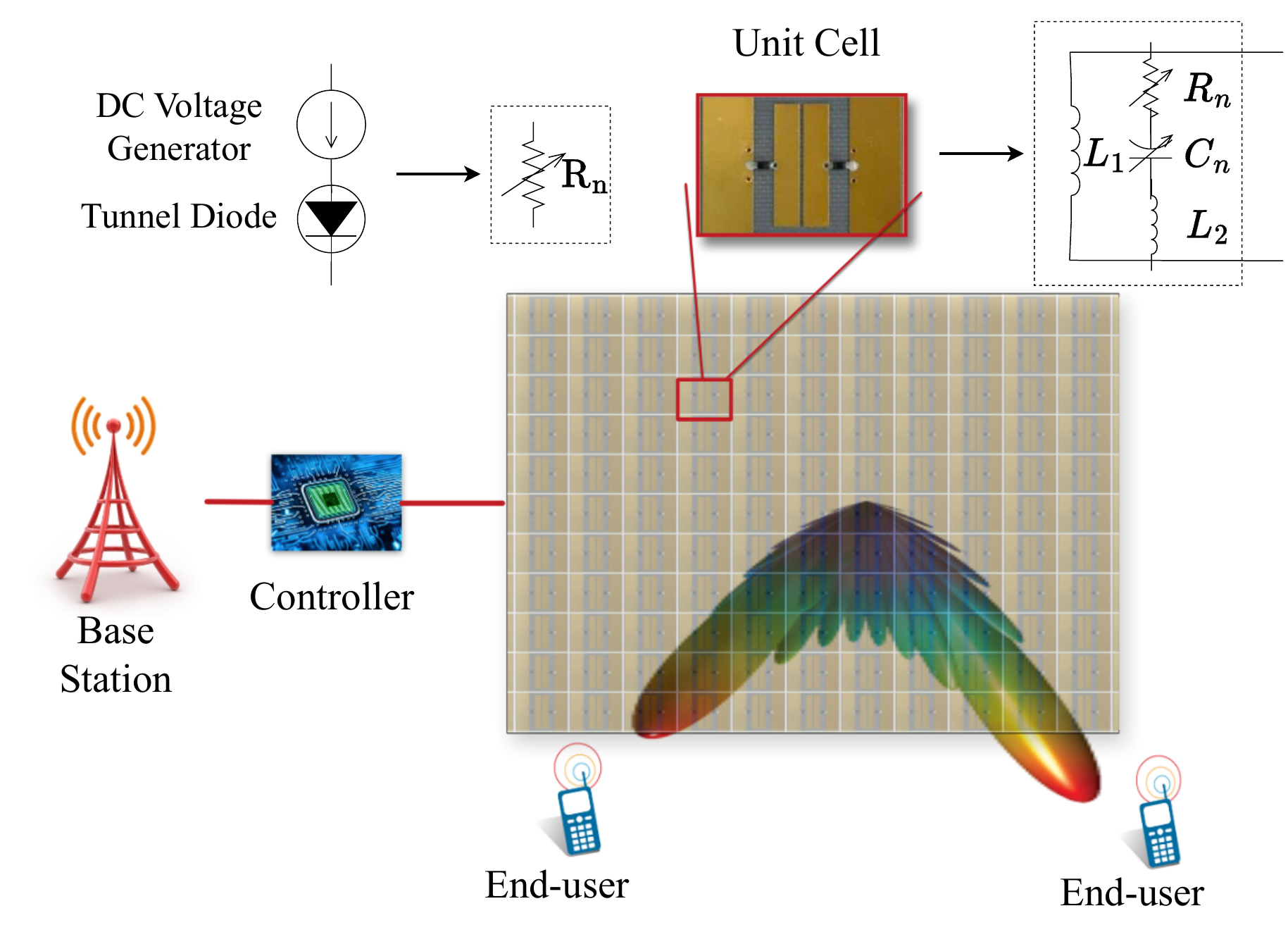}
    \caption{Illustration of an active RIS with TDs embedded in the transmission line circuit model of~\cite{Abeywickrama_2020}. Top-left zoom-in: The resistance \(R_n\) of the \(n\)-th element is replaced by a TD, controlled via a DC voltage source to enable reconfigurable amplitude tuning. Top-right zoom-in: The transmission line circuit model of the \(n\)-th unit cell. A centralized control architecture is assumed for both the RIS and the BS.}
    \label{fig:RIS_atom}
\end{figure}

However, integrating active elements requires careful attention to system stability, since uncontrolled negative resistance can lead to self-oscillations or even destructive feedback loops. To ensure stable amplification, the design must adhere to the stability criterion outlined in~\cite{green}, which requires the real part of the total impedance \(Z_n(C_n, R_n) + Z_0\) to remain strictly positive, while a nonzero imaginary part is necessary to avoid unwanted oscillations. Building upon the TD-based reflection amplifier architecture proposed in~\cite{Besma_full_duplex_TD}, where a full-duplex operation is achieved using tunnel diodes, this concept is hereinafter extended by embedding a TD within each RIS unit cell, as illustrated in Fig.~\ref{fig:RIS_atom}. In this configuration, the resistive component \(R_n\) is effectively replaced with a voltage-controlled TD that introduces a tunable negative resistance, thereby enabling reflection gain control via external DC biasing.

It is important to emphasize that real-world negative resistance devices, including TDs, exhibit a finite dynamic range for stable operation. Their amplification capability is bounded by saturation effects, and excessive input power can push the device outside its linear operating region, leading to distortion or instability~\cite{Besma_full_duplex_TD}. Nevertheless, in typical RIS applications, the received signal power at each unit cell is relatively low, ensuring that the device operates well within its linear regime. Thus, the assumption of linear amplification is practically justified and sufficient for the intended beamforming and reflection control tasks of the RIS.

\subsubsection{TDs: I-V Characteristic and Power Consumption}
\begin{figure}[t]
    \centering
    \includegraphics[width=0.8\columnwidth]{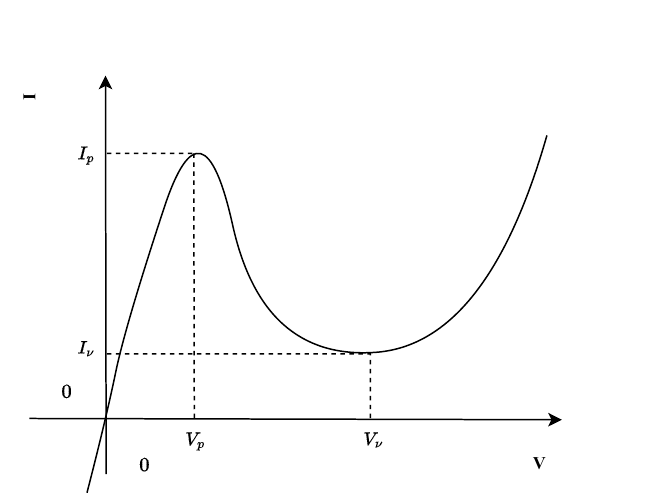}
    \caption{I-V characteristic curve of a typical TD, showcasing the peak ($V_p, I_p$) and valley ($V_v, I_v$) points. The negative resistance region lies between $V_p$ and $V_v$.}
    \label{fig:I-V_characteristic}
\end{figure}

As previously discussed, TDs exhibit a distinctive I-V relationship, characterized by several key regions \cite{sze_semiconductor}:
\begin{itemize}
    \item \textbf{Tunneling Region}: For applied voltages $V$ less than the peak voltage $V_p$, the current $I_T(V)$ increases with voltage, reaching a maximum at $V_p$. This behavior results from quantum tunneling, where charge carriers traverse the thin depletion region due to heavy doping levels in the diode's construction.
    \item \textbf{Negative Resistance Region}: Between the peak voltage $V_p$ and the valley voltage $V_v$, the diode exhibits negative differential resistance; that is, an increase in voltage leads to a decrease in current. This unique property enables applications in high-frequency oscillators and amplifiers.
    \item \textbf{Diffusion Region}: For voltages exceeding $V_v$, the current $I_d(V)$ increases again with voltage, resembling the behavior of conventional diodes.
\end{itemize}
A typical I-V characteristic curve of a tunnel diode is illustrated in Fig.~\ref{fig:I-V_characteristic}, highlighting the peak current $I_p$ at $V_p$ and the valley current $I_v$ at $V_v$.

In the following analysis, focus is given on the tunneling region where the cumulative current $I(V)$ is predominantly $I_T(V)$. According to semiconductor theory \cite{sze_semiconductor}, this tunneling current can be expressed as follows:
\begin{equation}\label{eq:I_V modeling}
    I_T(V) = \frac{I_p}{V_p} V \exp\left(1 - \frac{V}{V_p}\right).
\end{equation}
For a given set of parameters $V_p$, $I_p$, $V_v$, and $I_v$, the TD operates stably at the point in the negative resistance region where the resistance is minimized in magnitude. This holds because the derivative of the resistance with respect to the applied voltage is zero at that point, which ensures resilience against minor variations in bias voltage; this stable operating point is henceforth denoted by $R_{\rm sp}$.

To achieve different negative resistance values, one can modify the diode's intrinsic parameters. Factors influencing $V_p$ and $I_p$ include temperature, doping concentrations of the p-n junction, effective tunneling mass, and bandgap energy. For instance, increasing the doping levels sharpens the I-V characteristic in the tunneling region, leading to higher absolute values of negative resistance. For a more compact and parametric representation of the TD behavior, and to facilitate analysis of active RIS performance under varying negative resistance values, the generalized tunneling current model reported in~\cite{modelling_patent} is adopted:
\begin{equation}\label{eq:I_V modeling with m}
    I_T(V) = \frac{V}{R_0} \exp\left(-\left(\frac{V}{V_0}\right)^m\right),
\end{equation}
where \( R_0 \) denotes the Ohmic resistance in the diode's linear regime (i.e., for \( V < V_p \)), \( V_0 \in [0.1, 0.5] \) in Volts is a voltage scaling parameter, and \( m \in [1, 3] \) controls the steepness of the tunneling region. It is worth noting that \( R_0 \) can be interpreted as \( V_p/I_p \), thereby connecting this generalized model with~\eqref{eq:I_V modeling}.

The corresponding negative resistance $R(V)$ as a function of th applied voltage is given by:
\begin{equation}
    R(V) = \left( \frac{d I_T(V)}{dV} \right)^{-1} = R_0 \exp\left(\left(\frac{V}{V_0}\right)^m\right) \left(1 - m \left(\frac{V}{V_0}\right)^m\right)^{-1}.
\end{equation}
The stable operating point $R_{\rm sp}$ is then determined by solving $\frac{d R(V)}{d V} = 0$, yielding $V_{r} = (m^{-1} + 1)^{1/m} V_0$. Substituting $V_{r}$ into the expression for $R(V)$ yields:
\begin{equation}\label{eq:R_n}
    R_{\rm sp} = - R_0 m^{-1} \exp\left(\frac{m+1}{m}\right).
\end{equation}
The feasible negative resistance values $R_n$ for the TD-based active RIS unit elements are computed using this expression, with $m$ ranging from $1$ to $3$ and a specified $R_0$.

The power consumption $P_n$ required to configure the previously modeled TD-based negative resistance of each \(n\)-th active RIS unit element can be computed as follows. Recall that the focus here is on the power required by the active load, i.e., the TD operating in the negative resistance regime. When \( R_n < 0 \), the TD draws power due to the applied DC bias required to sustain the negative resistance. The consumed power in this regime is given using the voltage \(V_r\) at the stable operating point, as shown in~\cite{gavriilidis2025activeRIS}:
\begin{eqnarray}\label{eq:Pn_m}
P(R_n) \triangleq I_T(V_{ r})V_{ r} = \frac{V_0^2}{R_0}(m^{-1} + 1)^{2/m}.
\end{eqnarray}
which is a function of \(R_n\) due to \eqref{eq:R_n}. It can be seen from \eqref{eq:Pn_m} that the power consumption of each active RIS element depends on the tunneling current model parameters $V_0$, $R_0$, and $m$. However, recall from~\eqref{eq:R_n} that we assumed $R_0$ being fixed, since it models the Ohmic resistance before the tunneling region, hence, the variability of the negative resistance $R_n$ depends only on \(m \in [1, 3]\). Thus, it can be concluded that, by also fixing $V_0$, $P(R_n)$ as well as \(R_n\) will depend solely on the choice of \(m\).

It is important to emphasize that this model does not include the power consumption associated with configuring the reactive component \( C_n \). Although \( C_n \) is reconfigurable (e.g., via varactor diodes), the control circuitry typically consumes negligible power, especially in comparison to the power drawn by the active load. This approximation aligns with standard practice in the literature, where RIS elements are generally regarded as passive even when they support tunable phase shifts.

\subsubsection{Phase-Amplitude Dependence Model}
The reflection amplitude and phase shift of an active RIS element are not independently tunable; instead, they exhibit a structured interdependence governed by the element’s circuit-level parameters. In the context of the transmission-line circuit model introduced earlier, this relationship emerges from the combined effect of the tunable resistance \( R_n \) and capacitance \( C_n \) on the complex reflection coefficient \([\boldsymbol{\gamma}]_n(C_n, R_n)\) of each \(n\)-th RIS unit cell.

To formalize this dependency, the reflection coefficient iss expressed as \([\boldsymbol{\gamma}]_n(\alpha_n, \varphi_n) = \alpha_n e^{\jmath \varphi_n}\), where both the amplitude \(\alpha_n\) and phase shift \(\varphi_n\) are functions of the circuit parameters \( R_n \) and \( C_n \). Based on this formulation, the achievable amplitude range for a given phase shift is fist characterized. This is accomplished by solving~\eqref{eq:ref} with respect to \(\alpha_n\), yielding:
\begin{eqnarray}
\alpha_{\min}(\varphi_n) &\triangleq&  \sqrt{\frac{Z^{\rm I}_{n,\max} + (Z_0 - Z^{\rm R}_{n,\max}) \tan(\varphi_n)}{Z^{\rm I}_{n,\max} + (Z_0 + Z^{\rm R}_{n,\max}) \tan(\varphi_n)}}, \label{eq:amplitude_min}\\
\alpha_{\max}(\varphi_n) &\triangleq& \sqrt{\frac{Z^{\rm I}_{n,\min} + (Z_0 - Z^{\rm R}_{n,\min}) \tan(\varphi_n)}{Z^{\rm I}_{n,\min} + (Z_0 + Z^{\rm R}_{n,\min}) \tan(\varphi_n)}}, \label{eq:amplitude_max}
\end{eqnarray}
where \( Z_{n,\min} \triangleq Z_n(C_n(R_{\max}, \varphi_n), R_{\max}) \) and \( Z_{n,\max} \triangleq Z_n(C_n(R_{\min}, \varphi_n), R_{\min}) \), with \( R_{\min} \) and \( R_{\max} \) denoting the bounds of the tunable resistance range. The superscripts \({\rm R}\) and \({\rm I}\) denote the real and imaginary parts of the corresponding impedance values. Moreover, the capacitance \( C_n(R_n, \varphi_n) \) is itself a function of both the resistance and the desired phase shift, as derived in~\cite[eq.~13]{gavriilidis2025activeRIS}. Importantly, the feasible resistance range is phase-dependent; that is, \( R_{\min} \) and \( R_{\max} \) are more accurately represented as \( R_{\min}(\varphi_n) \) and \( R_{\max}(\varphi_n) \). This phase dependency arises because, increasing the magnitude of \( R_n \), constrains the range of realizable phase shifts. Conversely, targeting a specific \(\varphi_n\), restricts the admissible resistance values. This feasibility condition is investigated based on~\cite[eq.~13]{gavriilidis2025activeRIS}, by imposing the condition that \( C_n(R_n, \varphi_n) \) is considered valid only when it yields a real-valued capacitance. Using \eqref{eq:amplitude_min} and \eqref{eq:amplitude_max}, the amplitude for each RIS element can be expressed for \(\bar{\alpha}_n\in[0,1]\) as follows:  
\begin{equation}\label{eq:amplitude_wrt_phase}
    \alpha_n = \alpha_{\min}(\varphi_n) + \bar{\alpha}_n (\alpha_{\max}(\varphi_n) -\alpha_{\min}(\varphi_n) ).
\end{equation}

In the latter expression, \(\alpha_{\rm max}(\varphi_n)\) and \(\alpha_{\rm min}(\varphi_n)\) can be approximated using a linear steepness model (\(k=1\) in \cite[eq. (5)]{Abeywickrama_2020}), eliminating circuit parameter dependence. Extending~\cite{Abeywickrama_2020} to active RIS, we approximate the amplitude boundaries as follows:s
\begin{align}
\alpha_{\rm min}(\varphi) &\approx \frac{\delta_{\max}-\delta_{\min}}{2} ( \cos(\varphi + \theta) + 1 ) + \delta_{\min}, \label{eq:amplitude_min_approx}\\
\alpha_{\rm max}(\varphi) &\approx \frac{\beta_{\max}-\beta_{\min}}{2} ( \cos(\varphi + \theta) + 1 ) + \beta_{\min}, \label{eq:amplitude_max_approx}
\end{align}
where \(\delta_{\max} \triangleq \max_{\varphi} \alpha_{\min}(\varphi),\,\delta_{\min} \triangleq \min_{\varphi} \alpha_{\min}(\varphi),\,\beta_{\max} \triangleq \max_{\varphi} \alpha_{\max}(\varphi),\) and \(\beta_{\min} \triangleq \min_{\varphi} \alpha_{\max}(\varphi)\) are fitting variables that capture the effect of the negative resistance on the reflection coefficient's amplitude, and \(\theta\) is the phase shift where \(\delta_{\max}\) and \(\beta_{\max}\) occur \cite[eqs.~16 and 17]{gavriilidis2025activeRIS}.

The approximate expressions in~\eqref{eq:amplitude_min_approx} and \eqref{eq:amplitude_max_approx} allow to represent the RIS reflection coefficient vector in a form that explicitly captures the inherent phase-amplitude dependency using linear algebra. Specifically, by substituting these approximations into the expression \([\boldsymbol{\gamma}]_n = \alpha_n e^{\jmath \varphi_n}\), the reflection vector can be approximated as~\cite[eq.~20]{gavriilidis2025activeRIS}:
\begin{align}\label{eq:RIS_reflection_coefficient_vector}
    \boldsymbol{\gamma} \approx& \underbrace{\diag\left\{0.25 e^{\jmath \boldsymbol{\theta}} \odot (\mathbf{y} + \mathbf{x} \odot \boldsymbol{\bar{\alpha}}) \right\}}_{\triangleq \mathbf{Z}_2} \boldsymbol{\mathcal{D}}^{\tran}(\boldsymbol{\phi} \otimes \boldsymbol{\phi}) \nonumber\\
    & + \underbrace{\diag\left\{0.5 \mathbf{y} + \boldsymbol{\delta}_{\min} + (0.5 \mathbf{x} + \boldsymbol{\beta}_{\min} - \boldsymbol{\delta}_{\min}) \odot \boldsymbol{\bar{\alpha}} \right\}}_{\triangleq \mathbf{Z}_1} \boldsymbol{\phi} \nonumber\\
    & + \underbrace{0.25 e^{-\jmath \boldsymbol{\theta}} \odot (\mathbf{y} + \mathbf{x} \odot \boldsymbol{\bar{\alpha}})}_{\triangleq \mathbf{z}} \in \mathbb{C}^{N \times 1},
\end{align}
where the phase vector is defined as \(\boldsymbol{\phi} \triangleq [e^{\jmath \varphi_1}, \dots, e^{\jmath \varphi_N}]^{\tran} \in \mathcal{S}^N\), with \(\mathcal{S} \triangleq \{e^{\jmath \varphi} \mid \varphi \in [0, 2\pi]\}\), and
\begin{align*}
    \boldsymbol{\bar{\alpha}} &\triangleq [\bar{\alpha}_1, \dots, \bar{\alpha}_N] \in [0,1]^N, \\
    \mathbf{x} &\triangleq \boldsymbol{\beta}_{\max} - \boldsymbol{\beta}_{\min} - (\boldsymbol{\delta}_{\max} - \boldsymbol{\delta}_{\min}), \\
    \mathbf{y} &\triangleq \boldsymbol{\delta}_{\max} - \boldsymbol{\delta}_{\min}, \\
    \boldsymbol{\mathcal{D}} &\triangleq \left[\mathbf{D}_1 \, \mathbf{D}_2 \, \cdots \, \mathbf{D}_N\right]^{\rm T} \in \mathbb{B}^{N^2 \times N}.
\end{align*}
Here, each matrix \(\mathbf{D}_i \in \mathbb{B}^{N \times N}\) is a binary matrix with a single non-zero element at its \(i\)-th diagonal position, and \(\mathbb{B} \triangleq \{0,1\}\). The parameters \(\boldsymbol{\delta}_{\min}, \boldsymbol{\delta}_{\max}, \boldsymbol{\beta}_{\min}, \boldsymbol{\beta}_{\max},\) and \(\boldsymbol{\theta}\) are typically identical across all RIS elements and can be treated as constants.

\begin{figure}[!t]
 \includegraphics[width=\columnwidth]{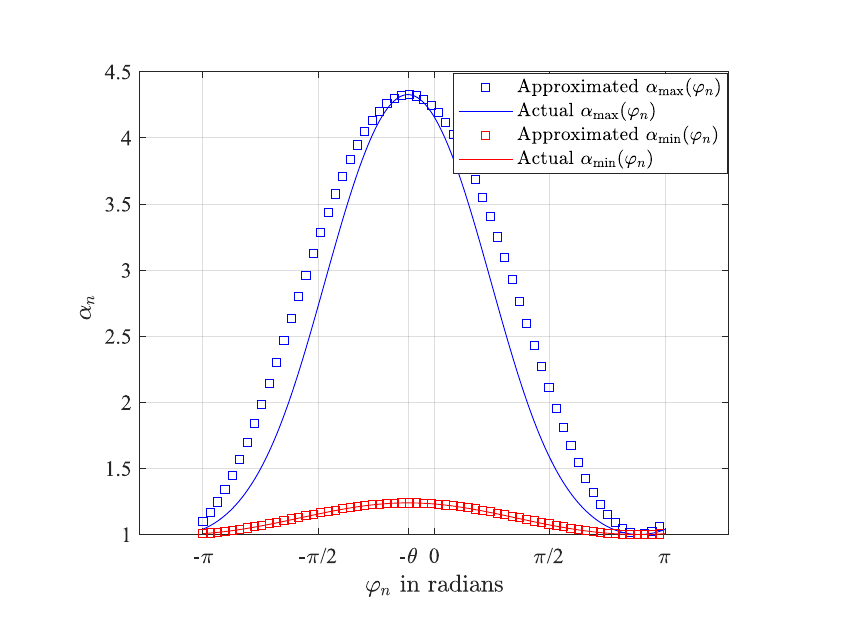}
 \caption{The upper, \(\alpha_{\max}(\varphi_n)\), and lower, \(\alpha_{\min}(\varphi_n)\), bounds for the amplitude for each \(n\)-th RIS active unit element as a function of the tunable phase value \(\varphi_n\), according to the active transmission line circuit model. The derived analytical formulas in \eqref{eq:amplitude_max} and \eqref{eq:amplitude_min} are compared with their respective approximations~\eqref{eq:amplitude_max_approx} and \eqref{eq:amplitude_min_approx}, considering $L_1$ = 4.5 nH, $L_2$ = 0.7 nH, $Z_0 = 377 \, \Omega$, $\omega = 2 \pi \times 2.4$ GHz, \(R_0 = 1\,\Omega\), \(R_n\equiv R_{\rm sp} \in [-7.39,-1.26]\)~$\Omega$ in~\eqref{eq:R_n}, and \(C_n \in [0.85,6.25]\) pF.}
 \label{fig:phase_amplitude_approximation}
 \end{figure}
The key benefit of the representation in~\eqref{eq:RIS_reflection_coefficient_vector} is that it recasts the reflection vector \(\boldsymbol{\gamma}\) in terms of two tunable vectors: the phase vector \(\boldsymbol{\phi}\) and the normalized amplitude vector \(\boldsymbol{\bar{\alpha}}\), where each \(\bar{\alpha}_n \in [0,1]\). This eliminates the need to directly manipulate the circuit-level parameters \(R_n\) and \(C_n\), which would otherwise result in a cumbersome and intractable formulation. More importantly, the intrinsic coupling between amplitude and phase is captured through a sequence of simple linear algebraic operations. This structure is particularly advantageous for optimization tasks, as it enables the use of convex relaxations and gradient-based algorithms for RIS configuration. Such methods would be computationally prohibitive if operating directly on the circuit parameters due to their complex, nonlinear dependency in the reflection model.

In Fig.~\ref{fig:phase_amplitude_approximation}, the exact expressions for the amplitude bounds (i.e., \eqref{eq:amplitude_min} and~\eqref{eq:amplitude_max}) are compared with their approximate counterparts (i.e., \eqref{eq:amplitude_min_approx} and~\eqref{eq:amplitude_max_approx}), considering a representative RIS element configuration. The results demonstrate that the approximations effectively capture the behavior of the exact expressions. In particular, the approximation of \(\alpha_{\max}(\varphi_n)\) shows minor deviations but remains highly accurate at the regions close the maximum and minima. On the other hand, the approximation for \(\alpha_{\min}(\varphi_n)\) exhibits an almost perfect fit across the entire phase range. These findings validate the accuracy of the presented linear steepness model in describing the phase-dependent behavior of the reflection amplitude, while significantly reducing the analytical complexity of the formulation.

\subsubsection{System Model and Rate Maximization}
Let us consider a communication system where a Tx with \(M_{\rm T}\) antennas aims to communicate with an Rxs having \(M_{\rm R}\) antennas. The Tx transmits \(d \leq \min\{M_{\rm T}, M_{\rm R}\}\) data streams, denoted by \(\mathbf{s} \in \mathbb{C}^{d \times 1}\), using spatial precoding represented by the matrix \(\mathbf{V} \in \mathbb{C}^{M_{\rm T} \times d}\), with the constraint \({\rm Tr}\{\mathbf{V}^H\mathbf{V}\} \leq P_{\rm T}\), where \(P_{\rm T}\) is the total transmit power budget. The communication is supported by an RIS with \(N\) TD-based active elements. The baseband received signal at the RX antenna elements is given by \cite[eq.~(1)]{energyeff}, augmented to account for the noise contributed by the RIS's active components, as follows:
\begin{align}\label{eq:system_model_antennas_streams}
  \mathbf{y} &\triangleq \underbrace{\left(\mathbf{H}_{\rm d} + \mathbf{H}_{2} \mathbf{\Gamma} \mathbf{H}_{1}\right)}_{\triangleq \tilde{\mathbf{H}}}
  \mathbf{V} \mathbf{s} + \mathbf{H}_2 \boldsymbol{\Gamma} \mathbf{n}_s + \mathbf{n}_r,
\end{align}
where \(\mathbf{H}_{\rm d} \in \mathbb{C}^{M_{\rm R} \times M_{\rm T}}\) represents the direct TX-RX channel, \(\mathbf{H}_1 \in \mathbb{C}^{N \times M_{\rm T}}\) is the RIS-TX channel, and \(\mathbf{H}_2 \in \mathbb{C}^{M_{\rm R} \times N}\) is the RX-RIS channel. The diagonal matrix \(\mathbf{\Gamma} \triangleq \mathrm{diag}\{\boldsymbol{\gamma}\} \in \mathbb{C}^{N \times N}\) models the RIS reflection coefficients, where \(\boldsymbol{\gamma} = \boldsymbol{\alpha} \odot \boldsymbol{\phi}\) with \(\boldsymbol{\alpha} \triangleq [\alpha_1, \ldots, \alpha_N]^{\rm T}\), and \(\alpha_n \in [\alpha_{\rm min}(\varphi_n), \alpha_{\rm max}(\varphi_n)]\). The noise terms \(\mathbf{n}_r \sim \mathcal{CN}(\mathbf{0}_{M_{\rm R} \times 1}, F_r \sigma^2 \mathbf{I}_{M_{\rm R}})\) and \(\mathbf{n}_s \sim \mathcal{CN}(\mathbf{0}_{N \times 1}, F_s \sigma^2 \mathbf{I}_N)\) represent the noise at the RX and RIS, respectively, where \(F_r\) and \(F_s\) denote the noise figures that characterize the SNR degradation due to the RIS and RX electronics. The received signal is processed by a combining matrix \(\mathbf{W} \in \mathbb{C}^{M_{\rm R} \times d}\), which leads to the detection of \(\mathbf{s}\) as \(\mathbf{\hat{s}} = \mathbf{W}^H \mathbf{y} \in \mathbb{C}^d\).

Given the knowledge of the channels \(\mathbf{H}_{\rm d}\), \(\mathbf{H}_1\), and \(\mathbf{H}_2\) (e.g., via any of the techniques described in~\cite{CE_overview_2022}), the goal hereinafter is to jointly optimize the TX precoder \(\mathbf{V}\), the RX combiner \(\mathbf{W}\), and the RIS phase-amplitude configuration \(\boldsymbol{\gamma}\) to maximize the achievable end-to-end MIMO rate. Assuming \(\mathbf{s} \sim \mathcal{CN}(\mathbf{0}_{d \times 1}, \mathbf{I}_d)\) (i.e., independent Gaussian symbols), the instantaneous spectral efficiency can be expressed as follows~\cite[eq. (6.61)]{heath}:
\begin{align}
 &\!\!\!\mathcal{R}(\mathbf{V},\mathbf{W},\boldsymbol{\gamma}) \!\triangleq \!\sum_{i=1}^d \!\log_2 \!\left(\!1 \!+\! \frac{||\mathbf{w}_i^H \tilde{\mathbf{H}} \mathbf{v}_i||_{\rm F}^2}{\!\sum \limits_{j=1, j \neq i}^d \!||{\mathbf{w}_i^H \tilde{\mathbf{H}} \mathbf{v}_j}||_{\rm F}^2 \!+ \!\sigma^2 F_s ||{\mathbf{w}_i \mathbf{H}_2 \mathbf{\Gamma}}||_{\rm F}^2\! +\! \sigma^2 F_r ||{\mathbf{w}_i}||_{\rm F}^2}\!\right)\!, \!\!\!\label{eq:R_RX_w_noise}
\end{align}
where \(||\cdot||_{\rm F}\) denotes the Frobenius norm.
Thus, the following optimization problem in favor of maximizing the spectral efficiency of the considered active-RIS-aided MIMO system can be formulated:
\begin{align}
\mathcal{OP}: &\max_{\mathbf{V}, \mathbf{W}, \boldsymbol{\gamma}} \mathcal{R}(\mathbf{V}, \mathbf{W}, \boldsymbol{\gamma}) \nonumber \\
& \,\,\,\, \text{s.t.} \,\, \text{Tr}\{\mathbf{V}^H \mathbf{V}\} \leq P_{\rm T}, \,\, \sum_{n=1}^N P(R_n) \leq P_{\rm RIS}, \label{OP_constraints_1} \\
& \,\,\,\, 0 \leq \varphi_n \leq 2\pi, \,\, \alpha_n \in [\alpha_{\min}(\varphi_n), \alpha_{\max}(\varphi_n)] \quad \forall n, \nonumber
\end{align}
where \(P_{\rm T}\) is the transmit power budget and \(P_{\rm RIS}\) denotes the total power available for the active RIS. To solve \(\mathcal{OP}\), the Alternating Optimization (AO) framework proposed in~\cite[Algorithm~1]{gavriilidis2025activeRIS} s adopted. First, the previous rate expression is reformulated by applying the optimal Linear Minimum Mean Squared Error (LMMSE) receive filter, which is given by:
\begin{equation}
\mathbf{W}_{\rm opt} = \left(\mathbf{\tilde{H}}\mathbf{V}\mathbf{V}^{\herm}\mathbf{\tilde{H}}^{\herm} \! +\! \sigma^2 F_s\, \mathbf{H}_2\mathbf{\Gamma}\mathbf{\Gamma}^{\herm}\mathbf{H}_2^{\herm} \! +\! \sigma^2 F_r\, \mathbf{I}_{M_{\rm R}}\right)^{-1} \mathbf{\tilde{H}}\mathbf{V}.
\end{equation}
The resulting rate expression becomes:
\begin{align}
    \mathcal{R}\left(\mathbf{V},\boldsymbol{\gamma}\right) =  &\sum_{i=1}^{d} \log_2\Biggl(1+ \mathbf{v}_i^{\herm}\mathbf{\tilde{H}}^{\herm}\biggl(\sum \limits_{j=1,\,j\neq i}^{d}\mathbf{\tilde{H}}\mathbf{v}_j\mathbf{v}_j^{\herm}\mathbf{\tilde{H}}^{\herm}+ \nonumber\\
     & \sigma^2 F_s\, \mathbf{H}_2\mathbf{\Gamma}\mathbf{\Gamma}^{\herm}\mathbf{H}_2^{\herm}+\sigma^2 F_r\, \mathbf{I}_{M_{\rm R}}\biggr)^{-1}\mathbf{\tilde{H}}\mathbf{v}_i\Biggr).\label{eq:Rate_linear_receiver}
\end{align}
Then, using the matrix quadratic and the Lagrangian dual transforms, \(\mathcal{OP}\) is reformulated as a convex problem with respect to \(\mathbf{V}\). The AO algorithm proceeds by iteratively updating the transmit precoder \(\mathbf{V}\), the RIS phase vector \(\boldsymbol{\phi}\) via manifold-based gradient ascent, and the amplitude vector \(\boldsymbol{\alpha}\) via convex optimization tools, until the spectral efficiency converges within a predefined threshold \(\epsilon\), or a maximum number of iterations \(J_{\rm alt}\) is reached.
At each iteration \(j\), the algorithm computes \(\mathbf{V}_{\rm opt}^{(j)}\), \(\boldsymbol{\phi}_{\rm opt}^{(j)}\), and \(\boldsymbol{\alpha}_{\rm opt}^{(j)}\), while the optimal combining matrix \(\mathbf{W}_{\rm opt}\) is directly obtained via the LMMSE expression above. Since the rate-maximizing receive filter depends only on the current \(\mathbf{V}\) and \(\boldsymbol{\gamma}\), and not vice versa, \(\mathbf{W}_{\rm opt}\) is computed once at the end of the iterative loop. To capture the phase-amplitude dependency during RIS phase optimization, the approximate reflection vector representation in~\eqref{eq:RIS_reflection_coefficient_vector} is utilized. Moreover, the RIS power constraint is linearized with respect to \(\boldsymbol{\alpha}\) using a Successive Convex Approximation (SCA) method during amplitude optimization. In the numerical results that follow, this AO-based solution is used to study the behavior of active-RIS-aided MIMO systems under different system conditions, such as SNR, RIS noise contributions, and constraints on power or the number of active RIS unit elements.

\subsubsection{Numerical Results and Discussion}

Unless otherwise stated, the following system parameters were used in our simulations: \(M_{\rm T} = M_{\rm R} = d = 8\), \(N = 64\), carrier frequency \(\omega = 2\pi \times 2.4\)~GHz, and noise power \(\sigma^2 = -113.93\)~dBm, corresponding to thermal noise over a \(1\)~MHz bandwidth at room temperature of \(20^\circ\)C. The receive and RIS-side noise figures were set to \(F_r = 7\)~dB and \(F_s = 2\)~dB, respectively. The available power budgets were chosen as \(P_{\rm RIS} = 2.3\)~W and \(P_T = -12.75\)~dBm. The Tx-RIS, RIS-Rx, and Tx-Rx distances were set to \({\rm d_{RIS,Tx}} = 40\)~m, \({\rm d_{Rx,RIS}} = 4\)~m, and \({\rm d_{Tx,Rx}} = 40.2\)~m. In addition, the TD-sbased active RIS unit elements were configured with the following parameters: \(V_0 = 0.1\)~V, \(L_1 = 4.5\)~nH, \(L_2 = 0.7\)~nH, \(Z_0 = 377\)~\(\Omega\), \(R_0 = 1\)~\(\Omega\), \(R_n \in [-7.39,-1.26]\)~\(\Omega\), and \(C_n \in [0.85, 6.25]\)~pF. For this configuration, the maximum amplification factor across all phases was \(\max_{\phi} \alpha_{\max}(\phi) = 4.3\), indicating that a single RIS active element can provide an equivalent gain to approximately \(4\) passive elements. Hybrid RIS architectures comprising both active and passive elements were also considered, where the passive elements are modeled with an Ohmic resistance of \(1.5\)~\(\Omega\). To emphasize the contribution of the RIS to the system performance, the direct MIMO channel between the TX and RX, denoted \(\mathbf{H}_{\rm d}\), was assumed to be blocked. This blockage was modeled by assigning a significantly higher path loss exponent of \(5\) to \(\mathbf{H}_{\rm d}\), in contrast to a path loss exponent of \(2\) applied to the RIS-assisted links \(\mathbf{H}_1\) (TX-to-RIS) and \(\mathbf{H}_2\) (RIS-to-RX); all channels weere sampled from the Rayleigh fading distribution.

For all spectral efficiency results, performance metrics were averaged over \(200\) independent Monte Carlo realizations. We define the normalized SNR parameter \(\rho \triangleq P_T \, \text{P}_L / (\sigma^2 F_r)\), where \(\text{P}_L \triangleq \frac{\lambda^4}{16 \pi^2}({\rm d_{RIS,TX}}{\rm d_{RX,RIS}})^{-2}\) represents the path loss of the cascaded RIS-assisted link (without accounting for the direct link). This quantity \(\rho\) corresponds to the received SNR of a SISO link with unit-gain scalar channel, and is used to efficiently characterize SNR-dependent behavior by varying a single parameter. For the default parameter values, \(\rho = -30\)~dB. In all subsequent results, the maximum number of iterations in \cite[Algorithm~1]{gavriilidis2025activeRIS} was set to \(J_{\rm alt} = 8\).

In the simulation results that follow, the performance of the AO algorithm proposed in~\cite{gavriilidis2025activeRIS} was evaluated and compared against general-purpose optimization frameworks, namely Genetic Algorithm (GA) and Particle Swarm Optimization (PSO). The parameters of all three approaches were configured to ensure identical asymptotic complexity, allowing for a fair comparison. Additionally, a Phase-Amplitude Independent (PAI) benchmark is included, which follows the same optimization procedure as AO but does not use the approximate RIS reflection model of~\eqref{eq:RIS_reflection_coefficient_vector}. Instead, it treats the amplitude and phase of each reflection coefficient as independent optimization variables. For the GA and PSO approaches, the optimal LMMSE receiver was used for combining, and the optimization was performed jointly over the circuit parameters \( R_n, C_n \), \( n = 1, \dots, N \), and the precoding matrix \( \mathbf{V} \).
\begin{figure}[t]
    \centering
    \includegraphics[width=\columnwidth]{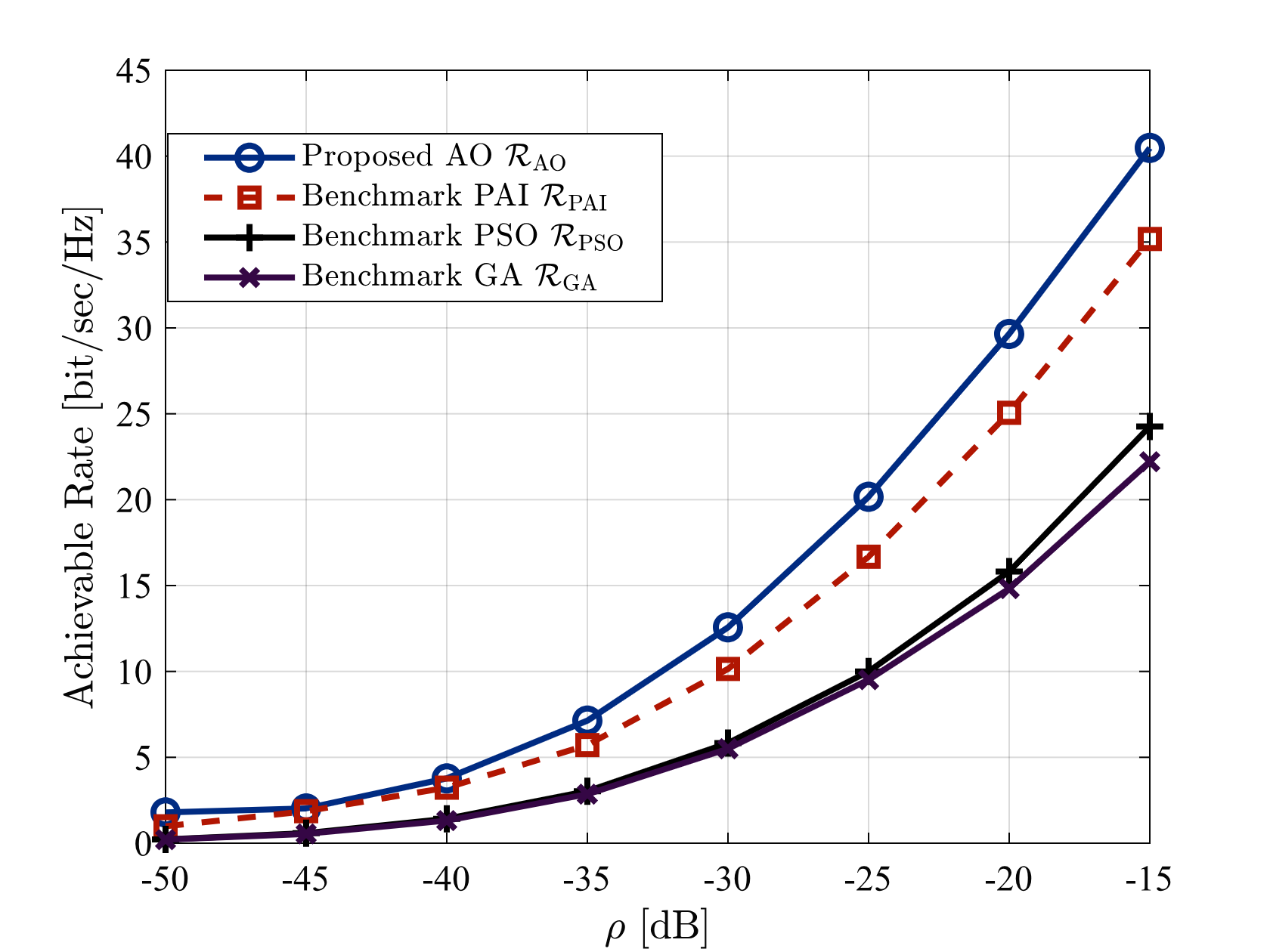}
    \caption{Achievable rate for varying \(\rho\) values, considering the AO algorithm proposed in \cite{gavriilidis2025activeRIS} and three benchmark schemes. For the PAI case, the AO framework without utilizing the phase-amplitude dependence formula \eqref{eq:RIS_reflection_coefficient_vector}, but treating \(\boldsymbol{\alpha}\) and \(\boldsymbol{\phi}\) as independent, was used.  
    }
    \label{fig: rate vs SNR}
\end{figure}

Figure~\ref{fig: rate vs SNR} depictes the achievable rate performance of the AO, PAI, GA, and PSO schemes as a function of the SNR variable \(\rho\), spanning from extremely low to high SNR regimes. As the \(\rho\) increases, a growing gap becomes evident between AO-based and the general-purpose methods, despite their equivalent complexity. This highlights the superior rate performance and convergence behavior of AO, confirming its effectiveness in capturing the structure of the problem. The comparison between AO and PAI further underscores the value of modeling the phase-amplitude dependency in the RIS reflection coefficients. While the performance difference is marginal in low-SNR conditions, where the rates are limited, the gap becomes increasingly pronounced at higher SNR values, indicating that neglecting the coupling between amplitude and phase can lead to significant suboptimality.
\begin{figure}[!t]
    \centering
    \includegraphics[width=\columnwidth]{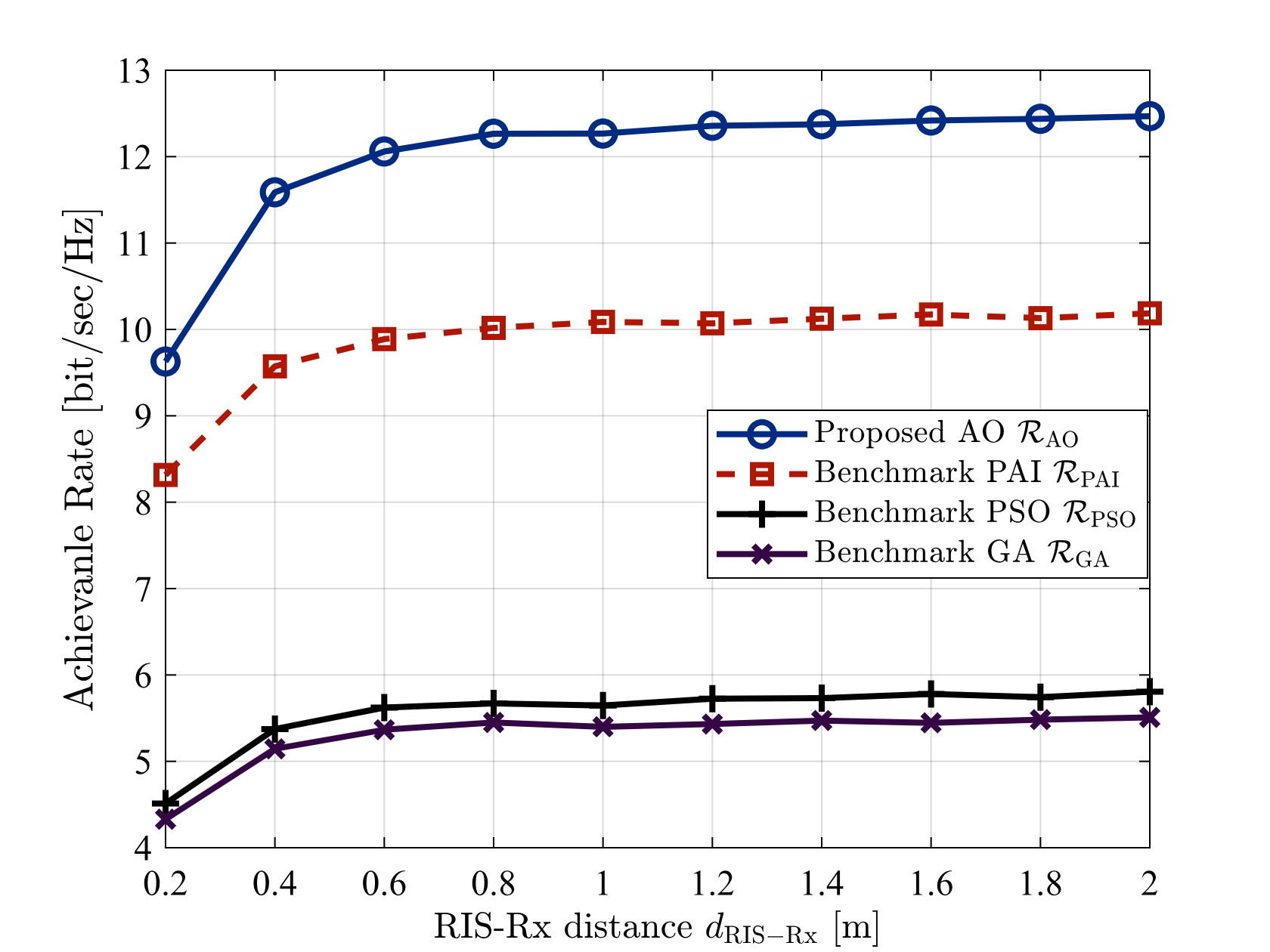}
    \caption{Achievable rate with respect to different \({\rm d_{RX,RIS}}\) values for \(\rho=-30\)~dB, considering three benchmark schemes. Decreasing \({\rm d_{RX,RIS}}\) increases the amplitude of \(\mathbf{H}_2\), thus, intensifying the noise contribution from the active RIS. This setup enabless to assess the performance of the active RIS across high- to low-power regimes of the interference noise.}
    \label{fig: rate vs distance}
\end{figure}

In Fig.~\ref{fig: rate vs distance}, the impact of RIS-generated noise is studied by varying the distance between the Rx and the RIS, denoted by \( d_{\rm Rx,RIS} \), while fixing \( \rho \) at  \(-30\)~dB. Essentially, the goal here is to investigate how the noise term generated from the RIS, which is given as \(\mathbf{H}_2\boldsymbol{\Gamma}\mathbf{n}_s\), impacts the rate performance, and, by varying the Rx-RIS distance \(d_{\rm RX,RIS}\), the channel \(\mathbf{H}_2\) is effectively attenuated or amplified, and consequently, the noise term. However, by keeping \(\rho\) fixed, which can be equivalently seen as changing the transmit power to counteract this path loss change, the intended signal power is kept the same for the cascaded link \(\mathbf{H}_2\boldsymbol{\Gamma}\mathbf{H}_1\mathbf{V}\mathbf{s}\), thus, allowing to isolate the effect of the RIS generated noise. The results show that, beyond a certain distance, the impact of this noise becomes negligible, and the performance of all schemes saturates. This observation provides a practical insight into when RIS-generated noise needs to be considered in system design. It can be seen that the AO framework remains the top-performing method, achieving more than twice the rate of PSO and GA, and outperforming the PAI benchmark by approximately \( 20\% \)–\( 25\% \).
\begin{figure}[!t]
    \centering
    \includegraphics[width=\columnwidth]{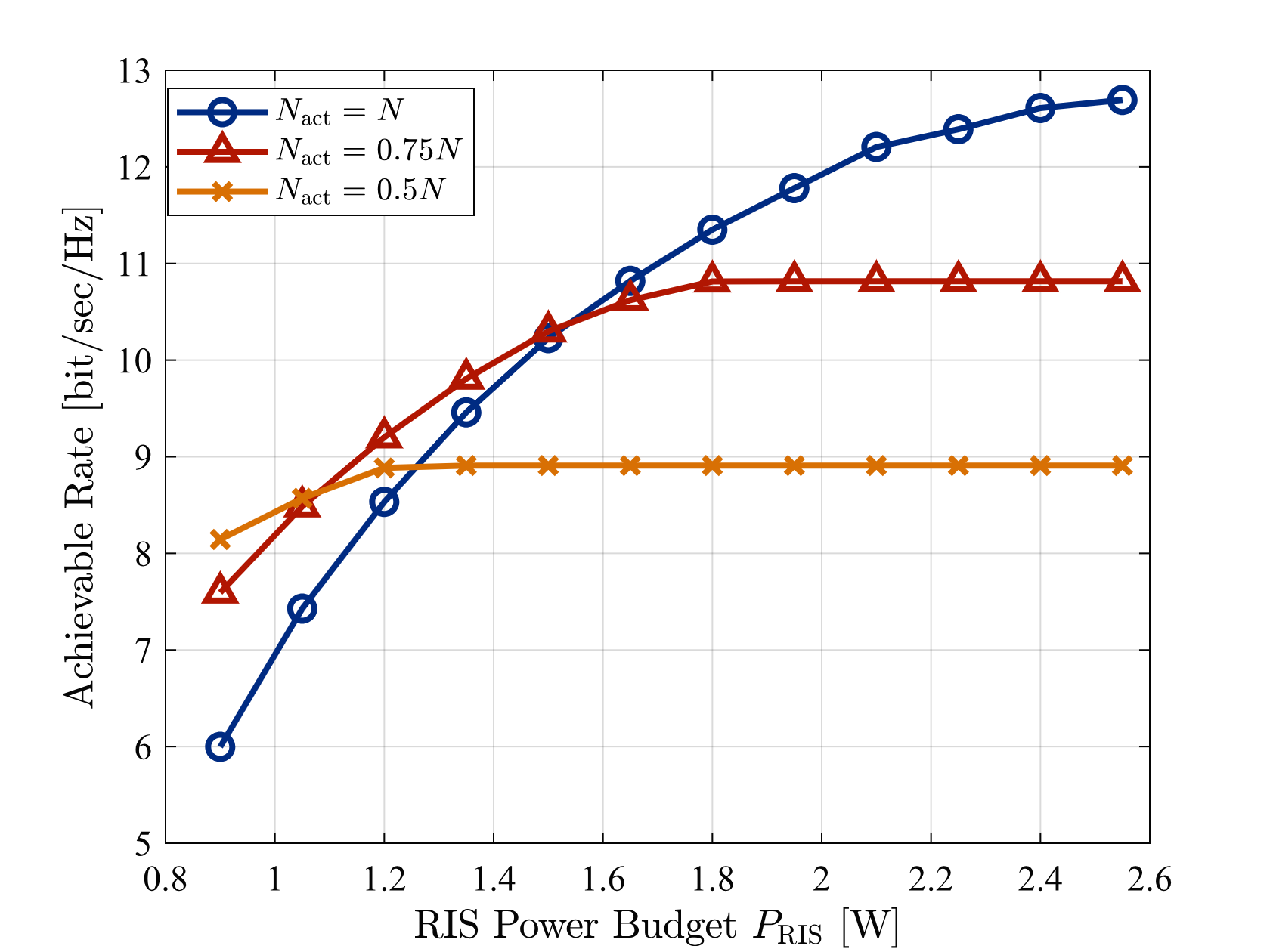}
    \caption{The achievable rate employing the AO framework versus varying RIS power budget constraints \(P_{\rm RIS}\). Different curves correspond to different number of active elements, considering that the RIS has a fixed number of $N=64$ mixed active and passive elements. }
    \label{fig: rate vs RIS power}
\end{figure}


To assess the performance of the active RIS under varying power budget constraints, Fig.~\ref{fig: rate vs RIS power} plots the achievable rate as a function of \( P_{\rm RIS} \) for \( P_{\rm RIS} \in [0.9, 2.55] \) W. Based on the system parameters defined earlier, the minimum and maximum power consumption of a single active element are \( P_{\min} = 12.1 \) mW and \( P_{\max} = 40 \) mW, respectively. For the considered \(64\)-element RIS, this translates to a total minimum power requirement of \( P_{\rm RIS} = 64 P_{\min} = 0.77 \) W for all elements to operate actively, and a maximum of \( P_{\rm RIS} = 64 P_{\max} = 2.56 \) W for full-power operation, which justifies the chosen range for \( P_{\rm RIS} \) in the simulation spanning from a tight power constraint (\( P_{\rm RIS} = 0.9 \) W) to a soft one (\( P_{\rm RIS} = 2.55 \) W). Interestingly, the results show that more active elements do not always improve performance; rather, the achievable rate strongly depends on the available power. Under a tight power budget, activating more elements results in inefficient power allocation and degraded performance. Conversely, activating fewer elements allows for greater individual amplification, often leading to higher rates. This trade-off is evident in the figure: for \( P_{\rm RIS} \leq 1.1 \) W, the configuration with \( N_{\rm act} = 0.5N \) yields the highest rate; in the intermediate range \( 1.1 \, \text{W} < P_{\rm RIS} \leq 1.5 \) W, the best-performing setup corresponds to \( N_{\rm act} = 0.75N \); and when the power budget is sufficiently large (\( P_{\rm RIS} > 1.5 \) W), it becomes optimal to activate all RIS elements. These results highlight the importance of optimizing the selection between active and passive elements based on the available power budget and the target rate performance, which remains an interesting direction for future research.

\section{Conclusions}\label{act_RIS_conc}
This chapter presented a comprehensive study on RISs with amplifying capabilities, offering detailed modeling frameworks, architectural designs, and performance analyses to address the limitations of traditional passive RISs. The discussion began with a comparison between passive and active RISs, highlighting the fundamental trade-offs between EE, cost, complexity, and achievable performance. While passive RISs offer low-power and low-cost implementations, it is already well known that their effectiveness is limited by the double path loss effect. Active RISs mitigate this issue by integrating amplifying components, enhancing system performance especially when the RIS is not placed near either of the communication end node. Hybrid configurations with some amplifying and passive elements and even single-amplifier designs emerge as promising solutions that balance performance with energy and hardware efficiency.

The chapter discussed two principal active RIS architectures. The first design, based on a single PA placed between two passive RIS panels, achieves amplification in the RF domain without requiring full baseband processing. A complete end-to-end system model was presented, including analytical expressions for the SNR and achievable rate. Using a Gamma distribution approximation for the SNR, the chapter also provided tractable expressions for the BEP under various system configurations. Numerical results demonstrated that this active design can effectively counteract the multiplicative path loss, but t is constrained by the PA's output power and gain limits. The EE analysis revealed that, despite higher power consumption, the single-PA active RIS often outperforms its passive counterpart with regards to EE due to the increased rate performance.

The second architecture considered TD-based reflection unit elements, leveraging the negative resistance region of the TD's I–V characteristic to enable amplification at the unit cell level in the RF domain, without requiring any baseband conversion. A detailed circuit-level model was presented that captures the interaction between the tunable resistance and capacitance values that govern each element's reflection coefficient. To enable tractable analysis and efficient optimization, an approximate linear steepness model was introduced, capturing the intrinsic phase-amplitude coupling without relying on explicit hardware-level parameters. Building on this model, a rate maximization problem was formulated for an RIS-assisted MIMO system, incorporating realistic circuit-level constraints, including transmit power limitations and the RIS’s total power budget. The impact of accounting for the phase-amplitude dependency in the optimization was highlighted, and several key insights were derived regarding the trade-off between the number of active elements and the available power budget.

In summary, this chapter established that RISs with amplifying capabilities, whether realized through centralized PAs or distributed negative resistance elements, can significantly expand the operational flexibility and performance potential compared to their passive counterparts. While practical considerations such as PA saturation, hardware complexity, and stability must be carefully managed, the presented designs and models in this chapter provide a solid foundation for future research and implementation of scalable, energy-aware RIS-assisted communication systems in 6G and beyond wireless systems.

\bibliographystyle{unsrt}
\bibliography{references}
\end{document}